\documentclass[10pt, aps, pra, amsmath, amssymb, amsfonts, floatfix, twocolumn, superscriptaddress]{revtex4-2}

\usepackage[T1]{fontenc}
\usepackage[utf8]{inputenc}
\usepackage{graphicx}
\usepackage{bm}
\usepackage[separate-uncertainty]{siunitx}
\usepackage[breaklinks=true]{hyperref}
\hypersetup{
colorlinks=true,
linkcolor=blue,
citecolor=blue,
urlcolor=blue
}
\usepackage[dvipsnames]{xcolor}

\newcommand{\sectionmajor}[1]{
\vspace{1em}
\noindent
{\large\textbf{#1}}
}
\newcommand{\sectionminor}[1]{
\vspace{1em}
\noindent
{\textbf{#1}}
}
\begin{document}

\title{Broadband coherent XUV light from {$e^-/e^+$} microbunching in an intense laser pulse}

\author{Michael~J.~Quin}
\email{michael.quin@physics.gu.se}
\affiliation{Max-Planck-Institut f\"{u}r Kernphysik, Saupfercheckweg 1, 69117 Heidelberg, Germany}
\affiliation{Department of Physics, University of Gothenburg, 41296 Gothenburg, Sweden}

\author{Antonino~Di~Piazza}
\email{a.dipiazza@rochester.edu}
\affiliation{Department of Physics and Astronomy, University of Rochester, Rochester, NY 14627, USA}
\affiliation{Laboratory for Laser Energetics, University of Rochester, Rochester, NY 14623, USA}
\affiliation{Max-Planck-Institut f\"{u}r Kernphysik, Saupfercheckweg 1, 69117 Heidelberg, Germany}

\author{\c{C}a\u{g}r{\i} Erciyes}
\affiliation{Max-Planck-Institut f\"{u}r Kernphysik, Saupfercheckweg 1, 69117 Heidelberg, Germany}

\author{Christoph~H.~Keitel}
\affiliation{Max-Planck-Institut f\"{u}r Kernphysik, Saupfercheckweg 1, 69117 Heidelberg, Germany}

\author{Matteo~Tamburini}
\email{matteo.tamburini@mpi-hd.mpg.de}
\affiliation{Max-Planck-Institut f\"{u}r Kernphysik, Saupfercheckweg 1, 69117 Heidelberg, Germany}

\date{\today}

\begin{abstract}
\noindent 
Attosecond pulses of coherent extreme ultraviolet (XUV) light are instrumental for investigating subatomic dynamics and can be produced using a free-electron laser (FEL). It has been suggested that an optical FEL, which employs a laser pulse in place of a conventional undulator, could enable a dramatically more compact implementation of such a light source. Yet, the high electron density and subsequent high emittance implied by an optical FEL makes this concept challenging to realize with an electron beam. There has been impressive progress in recent years producing collimated dense and relativistic beams of electrons and positrons in the laboratory. As we demonstrate here, the inherent stability of a quasi-neutral electron-positron beam mitigates Coulomb expansion, and renders it a promising alternative source of coherent light. Specifically, we show via computer simulations that broadband coherent light in the XUV domain, which takes the form of 8-$\si{\atto\second}$ pulses at 92-$\si{\atto\second}$ intervals, can be generated by microbunching of relativistic electrons and positrons in a laser pulse. This process occurs over a sub-millimeter length scale, enabling the development of light sources which are orders-of-magnitude more compact than existing sources, with potential applications in physics, chemistry, biology, and industry.

\end{abstract}

\maketitle

\sectionmajor{Introduction}
\smallskip

\noindent
Coherent pulses of extreme ultraviolet (XUV) light are capable of tracking electron dynamics within atoms and molecules, which take place on attosecond time scales~\cite{krausz2009, calegari2016}. Light sources of this kind have been employed  to study fundamental processes such as photoemission and tunnelling ionization in atomic, molecular and condensed matter systems~\cite{lepine2014, nisoli2017, ciappina2017, calegari2023}. Attosecond pulses are typically produced by high harmonic generation~\cite{hentschel2001}, and impressive progress has been made extending this mechanism into the soft x-ray domain~\cite{teichmann2016, gaumnitz2017, li2017}. Yet, the low conversion efficiency at high harmonic orders limits the intensity and applicability of the emitted light. This has prompted the generation of attosecond pulses from free electron lasers (FELs)~\cite{duris2020, maroju2020, franz2024}, which are far less compact, but rank among the brightest sources of x-rays available.

Coherent emission in a FEL takes place due to the formation of microbunches which are shorter than the wavelength of emitted radiation. Specifically, shorter than the first harmonic ${\lambda_{\text{FEL}}=\lambda_u(1+\frac{1}{2}K^2)/2\gamma^2_0}$ of radiation emitted on-axis by ultra-relativistic (Lorentz factor $\gamma_0\gg1$) electrons as they pass through an undulator with strength parameter $K=|e| B_0 \lambda_u / 2 \pi m$, magnetic field $B_0$ and spatial period $\lambda_u$~\cite{huang2007, schmueser2009}. Here, we employ natural units where $c=\hbar=4\pi\varepsilon_0=1$. We denote the particle's charge as $e$ and its mass as $m$, with $e<0$ for an electron. Microbunching can be induced by the self interaction of the electron bunch with its emitted radiation~\cite{emma2010} or by a seed laser~\cite{allaria2012}. In the latter case, attosecond-FELs typically compress the bunch in a magnetic chicane and utilize an infrared laser as a seed~\cite{duris2020, maroju2020, franz2024}. Broadband coherent radiation would significantly improve the efficiency of crystallography, emission and absorption spectroscopy, stimulated Raman spectroscopy, and multi-wavelength anomalous diffraction. For FELs, this can be achieved by increasing the chirp of the electron beam~\cite{prat2020}.

There are a couple of methods by which the size of an attosecond-FEL could be reduced. One is to substitute the linac with a compact, plasma-based accelerator~\cite{pompili2022, labat2023, habib2023}. Another, would be to replace the undulator entirely with an optical laser pulse  counter-propagating with respect to the electron beam. This is known as an optical-FEL (OFEL)~\cite{geabanacloche1987, gallardo1988, steiniger2014, steiniger2019}. Modelling the laser pulse as a linearly-polarized plane wave of amplitude $a_0=|e|E_0/m\omega_0$, electric field $E_0$ and central frequency $\omega_0=2\pi/\lambda_0$, the first harmonic emitted on-axis will be~\cite{sarachik1970, salamin1996}
\begin{equation}
	\lambda_1=\frac{\lambda_0}{4\gamma^2_0}\left(1+\frac{1}{2}a_0^2\right).
 \label{eq:first}
\end{equation}
This differs from $\lambda_{\text{FEL}}$ by one-half, as the phase depends on both the spatial coordinate and time in a laser pulse.
 
However, an optical wavelength $\lambda_0=800\,\si{\nano\metre}$ is far smaller than the spatial period $\lambda_u\approx 27\,\si{\milli\metre}$ of a typical undulator~\cite{schreiber2015}. For $\lambda_1 \sim \lambda_{\text{FEL}}$, this implies the production of dense and relatively low energy microbunches in an OFEL. As such, OFELs are challenging to realize because such low energy and high-density electron beams suffer from Coulomb repulsion and high emittance, hindering microbunch formation and coherent emission.

\begin{figure}
	\centering
	\includegraphics[width=0.99\linewidth]{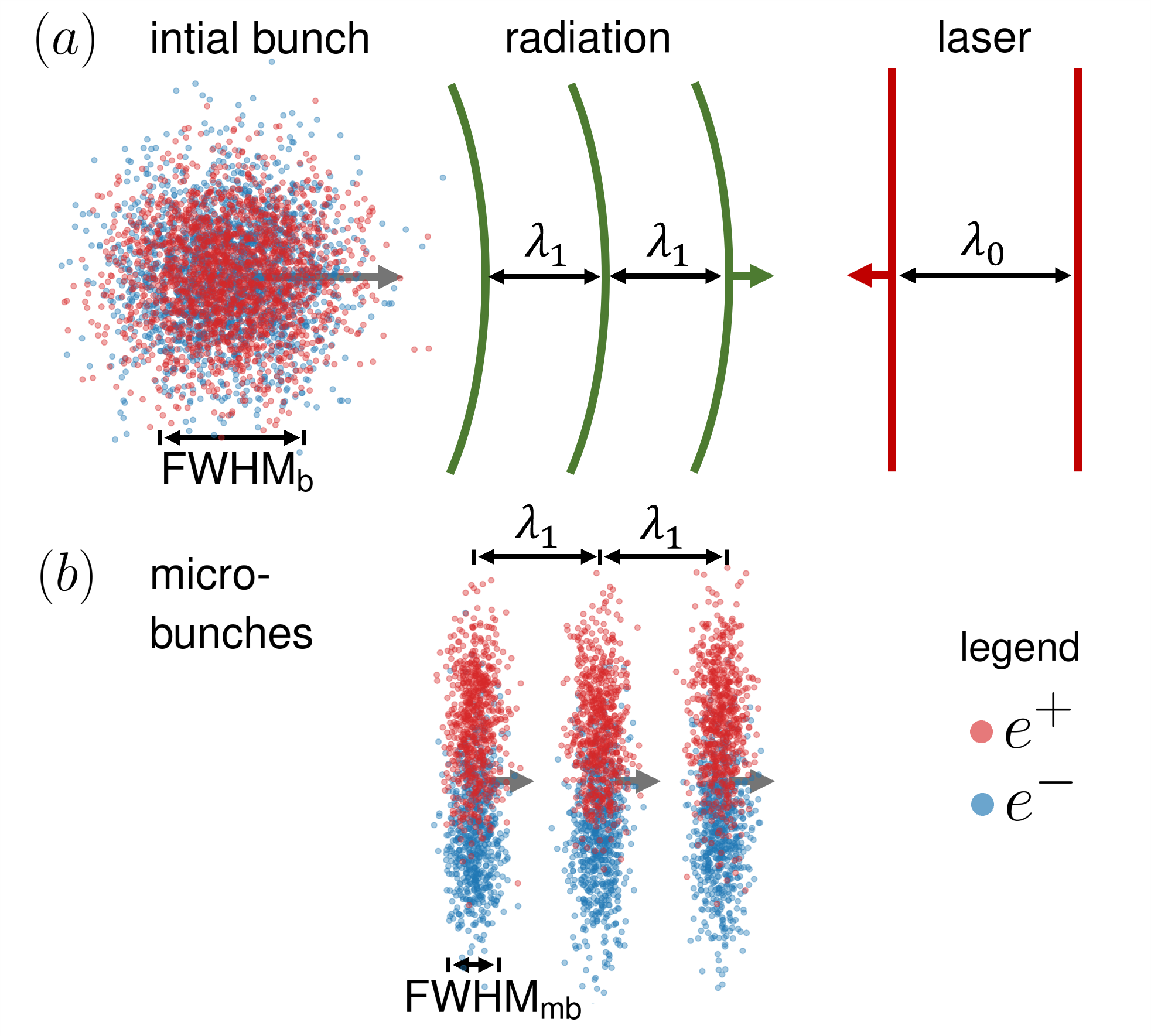}
	\caption{\textbf{Schematic of microbunching in a laser pulse.} (a) Collision of $e^-/e^+$ bunch (blue and red dots) with a plane wave of wavelength $\lambda_0$ (red lines) to produce harmonics of radiation starting from $\lambda_1$ (green curved lines). (b) Microbunching due to the interaction of the bunch with its self-generated radiation, at the laser pulse peak. Here $\lambda_1$ controls the microbunch separation and provides an upper limit of the width $\text{FWHM}_{\text{mb}}$. 
	}
	\label{fig:schematic}
\end{figure}

In this paper, we show how broadband coherent XUV light can be obtained by microbunching of electrons and positrons ($e^-/e^+$) in an intense laser pulse ($a_0^2\gg1$). A schematic of this process is shown in Fig.~\ref{fig:schematic}. Radiation emitted from this system takes the form of an attosecond pulse train with a well controlled spectral phase. Systems of electrons and positrons are inherently more stable, due to their equal mass and opposite charge. Specifically, the beam neutrality suppresses Coulomb expansion and provides a restoring force, which allows for sustained coherent emission. Therefore, a compact OFEL can be realized if a high quality $e^-/e^+$ beam is provided.

On this point, we note the steady improvement in the yield, density and energy of $e^-/e^+$ sources which has taken place in recent years~\cite{sarri2015, chen2015, liang2015, xu2016, peebles2021, arrowsmith2024, streeter2024, noh2024}. These experiments typically involve an ultra-relativistic beam of charged particles from a plasma-based accelerator or linac propagating through a high-Z target to generate $e^-/e^+$ pairs via the Bethe-Heitler process. We suggest that the scheme presented here could be realized in the near future by compressing a high quality $e^-/e^+$ beam from one of these sources in a magnetic chicane.

\sectionmajor{Results}

\vskip 2mm
\noindent
We begin by outlining the system of equations solved by our point particle code, which is employed to demonstrate the microbunching effect, radiation properties and stability of the system from first principles (see Methods for details on the code). Then, we perform fully three-dimensional particle-in-cell (PIC) simulations to simulate an $e^-/e^+$ beam with more realistic parameters, bringing the scheme closer to implementation.

\sectionminor{Dynamics of point particles.}
The causality-preserving solution of Maxwell's equations for a point-like charge is given by the Li\'{e}nard-Wiechert fields, which can be separated into `velocity' and `acceleration' fields~\cite{rohrlich2007}
\begin{align}
    F_{\text{LW}\,i}^{\mu\nu}(x) &= F_{\text{vel}\,i}^{\mu\nu}(x) + F_{\text{acc}\,i}^{\mu\nu}(x),
    \label{eq:Fret}
    \\%[0.1em]
    F_{\text{vel}\,i}^{\mu\nu}(x) &= \left[\frac{2e_i}{R^2_i}\,\frac{n^{[\mu}_i u^{\nu]}_i}{(n_iu_i)^3}\right]_{t_{\text{ret}\,i}} 
    \label{eq:Fvel},
    \\%[0.1em]
    F_{\text{acc}\,i}^{\mu\nu}(x) &= \left[\frac{2e_i}{R_i}\,\frac{n^{[\mu}_i \mathrm{a}^{\nu]}_i}{(n_iu_i)^2} - \frac{2e_i}{R_i}\,\frac{n^{[\mu}_i u^{\nu]}_i}{(n_iu_i)^3}(n_i\mathrm{a}_i)\right]_{t_{\text{ret}\,i}}. 
    \label{eq:Facc}
\end{align}
These describe the field seen at $x^\mu=(t, \bm{x})$ produced by a particle with four-velocity $u^\mu_i$ and four-acceleration $\mathrm{a}^\mu_i$, separated by distance $R_i=|\bm{x}-\bm{x}_i|$ in direction $n^\mu_i=(x^\mu-x_i^\mu)/R_i$ at the retarded time $t_{\text{ret}\,i}=t-R_i$. Here we use the Minkowski metric $\text{diag}(+1, -1, -1, -1)$ and short-hand notation $(ab) \equiv a^\mu b_\mu$ and $a^{[\mu} b^{\nu]} \equiv \frac{1}{2}(a^\mu b^\nu - a^\nu b^\mu)$. If the observer is another distinct particle $j\neq i$ then we refer to $F_{\text{LW}\,i}^{\mu\nu}(x_j)$ as an `interparticle' field and, in the particular case where both particles are of the same species, as an `intraspecies' field. 

The essential novelty of our point-particle code is its ability to describe the mutual interaction of structureless classical point particles and to distinguish between the velocity field, which represents a Coulomb-like interaction, and the acceleration field, which represents emitted radiation. Specifically in the $e^-/e^+$ system considered here, we will show that the acceleration field is responsible for microbunching, while the velocity field provides a restoring force between $e^-$ and $e^+$ which stabilizes the system and increases its coherence. We have checked numerically that the Coulomb divergence of the Li\'{e}nard-Wiechert fields has no practical consequences in our simulations.

\begin{figure*}[t]
    \centering
    \includegraphics[width=0.999\linewidth]{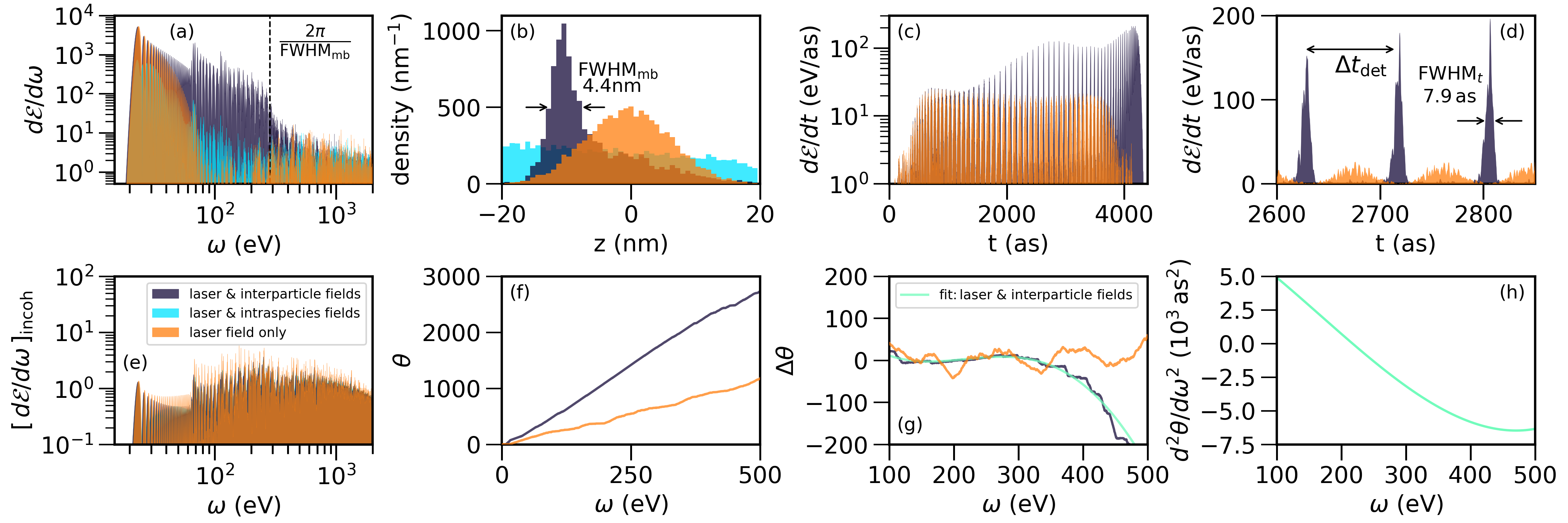}
    \caption{\textbf{Collision of $e^-/e^+$ bunch with laser pulse.} From simulations with our point particle code: (a) spectrum of energy radiated onto $1\,\si{\centi\metre\squared}$ detector at $1\,\si{\metre}$ distance along $+z$ from the collision region, with first harmonic $\omega_1\approx23\,\si{\electronvolt}$. (b) Bunch compression at laser pulse peak. (c) Attosecond pulse train observed at detector. (d) Properties of the attosecond pulses and their temporal separation $\Delta t_{\text{det}}\approx92\,\si{\atto\second}$. (e) Spectrum of incoherent energy radiated onto detector; field configurations are overlapping and indistinguishable. (f) Spectral phase of radiation. (g) Residual spectral phase. (h) Group delay dispersion evaluated from the polynomial fit in (g). Legends in (e) and (g) apply to all plots. The bunch, compressed by the interaction with its own radiation, produces coherent XUV light in the form of an attosecond pulse train with a stable spectral phase.
    }
    \label{fig:LP400nm}
\end{figure*}

The total field seen by each particle is then the sum of the external (laser) field and interparticle fields
\begin{equation}
	\mathcal{F}^{\mu\nu}_i(x_i) = F^{\mu\nu}_{\text{ext}}(x_i) + \sum^N_{\substack{j=1\\j\neq i}} F^{\mu\nu}_{\text{LW}\,j}(x_i).
	\label{eq:Ftot}
\end{equation}
Finally, the trajectories can be evaluated by solving the `reduced' Landau-Lifshitz equation~\cite{landaulifshitz_vol2}
\begin{equation}
	m_i\mathrm{a}^\mu_i = e_i \mathcal{F}^{\mu\nu}_i u_{\nu,i} + \frac{2e^4_i}{3m_i^2} \Big[\Big. \mathcal{F}^{\mu\nu}_i \mathcal{F}_{\nu\alpha,i} u^\alpha_i + (\mathcal{F}_iu_i)^2 u^\mu_i \Big.\Big],
	\label{eq:LLeq}
\end{equation}
which accounts for the energy and momentum loss during the emission of radiation via a self-force~\cite{rohrlich2007, dipiazza2012_review}. This effect, known as radiation reaction, is small for the parameters considered here, but is nevertheless included to ensure energy and momentum conservation~\cite{quin2023}. Here, `reduced' indicates that we have neglected derivatives of the field as they are typically much smaller than quantum-mechanical corrections~\cite{tamburini2010}, which are also negligible for the parameters considered here.

For a relativistic $e^-/e^+$ bunch, the classical description employed here is valid provided that the average interparticle distance is well above the Bohr radius of positronium in the average rest frame, preventing bound state formation and electron-positron annihilation. Additionally, we require that the field experienced by each particle in its instantaneous rest frame is small compared to the critical field of quantum electrodynamics $E_{\text{cr}}=m^2/|e|\approx1.3\times10^{18}$\,\si{\volt/\metre}, i.e., $\chi_0\approx2\gamma_0 E_0/E_{\text{cr}} \ll 1$~\cite{dipiazza2012_review, gonoskov2022, fedotov2023}. Both of these conditions are satisfied throughout our simulations. 

With the trajectories known, the spectrum of energy radiated ${d\mathcal{E}/d\omega d\Omega=(4\pi^2)^{-1}|\sum^N_{i=1}\bm{\mathcal{I}}_i(\omega, \bm{n})|^2}$ as seen by a distant observer along $n^\mu=(1,\bm{n})$, satisfying $(n)^2=0$, is obtained by the radiation integral~\cite{jackson1998}
\begin{equation}
    \bm{\mathcal{I}}_i(\omega, \bm{n}) = e_i\int^{+\infty}_{-\infty}\frac{d}{dt} \left[ \frac{\bm{n}\times(\bm{n}\times\bm{u}_i)}{(nu_i)}\right] e^{i\omega(nx_i)}\,dt.
    \label{eq:energy}
\end{equation}
The incoherent part of the spectrum is similarly defined as ${[d\mathcal{E}/d\omega d\Omega]_{\text{incoh}} = (4\pi^2)^{-1}\sum^N_{i=1}|\bm{\mathcal{I}}_i(\omega, \bm{n})|^2}$.

Details of the numerical code developed to solve this system of equations can be found in the Methods section. A key advantage over alternative codes is the separation of interparticle fields into acceleration and velocity components. By artificially `switching off' the velocity fields $F^{\mu\nu}_{\text{vel}\,j}(x_i)=0$ from all particles $j$, we can isolate the role of the radiation from the electrostatic-like attraction between different species of particles.

\sectionminor{Simulation setup.} All of our simulations consider the collision of a relativistic and neutral $e^-/e^+$ bunch with an intense ($a_0^2\gg 1$), linearly-polarized laser pulse. In this regime, a quasi-continuous series of harmonics is emitted on-axis beginning from $\omega_1=2\pi/\lambda_1$~\cite{sarachik1970, salamin1996}. Coherent emission at an angular frequency $\omega$ is expected providing the initial full-width-at-half-maximum FWHM$_{\text{b}}$ of the bunch satisfies $\omega\,\text{FWHM}_{\text{b}}\lesssim2\pi$. This condition automatically implies only a small energy spread and divergence can be tolerated, otherwise the bunch would quickly expand and coherent emission would cease.

In simulations with our point particle code, we consider a neutral Gaussian $e^-/e^+$ bunch of FWHM$_{\text{b}}=16\,\si{\nano\metre}$ containing 4000\,$e^-$ and 4000\,$e^+$, moving along $+z$ with a kinetic energy of $2.0\,\si{\mega\electronvolt}$ ($\gamma_0=5$), divergence $\sigma_\vartheta=1\,\si{\milli\radian}$, and kinetic energy spread $\sigma_{\text{KE}}=0.1\%$. Note that sub-permille energy spread has very recently been achieved via active energy compression of a laser-plasma-generated electron beam~\cite{winkler2025}. One can show the average interparticle distance in the rest frame $R\approx24\,a_{\text{ps}}$ is an order of magnitude above the Bohr radius for positronium $a_{\text{ps}}\approx0.11$\,\si{\nano\meter}, indicating that bound state formation and $e^-/e^+$ annihilation are unlikely~\cite{quin2023_phd}. This remains true throughout the simulation. 

The $e^-/e^+$ bunch collides head-on with a laser pulse, linearly polarized along $x$, propagating along $-z$, and modeled as a plane wave pulse. The plane wave approximation has no significant impact on our simulation results, as described in the Methods section, and allows the microbunching effect to be demonstrated with a simple analytical model. The laser pulse has a fixed amplitude $a_0=5$, pulse length $\text{FWHM}_L\approx26.7\,\si{\femto\second}$, central wavelength $\lambda_0=400~\si{\nano\metre}$ and cycle-averaged peak intensity $I_0\approx 2.2\,\times10^{20}\,\si{\watt/\centi\metre^2}$. This specific wavelength was not chosen for any physical reason, but rather to reduce the cost of our simulations. A laser pulse of this kind can be obtained by frequency-doubling a Ti:Sapphire laser~\cite{wang2017}, where the efficiency of this process can exceed 50\,\%~\cite{mironov2011}.

In order to unveil the physics of the microbunching process, we consider several configurations for the field $\mathcal{F}^{\mu\nu}_i(x_i)$ when solving the reduced Landau-Lifshitz equation: (i) `laser \& interparticle fields', where particles interact with the total field in Eq.~\eqref{eq:Ftot}, (ii) `laser \& intraspecies fields', where particles interact with the total field excluding fields from different species of particles, that is the $e^-$ and $e^+$ do not interact, and (iii) `laser field only', where particles interact with only the external laser field. In addition, we consider (iv) `laser \& acceleration fields', which is the same as (i) except the velocity fields have been artificially switched off in the simulation.

\sectionminor{Bunch compression via radiation emission.}
Figure~\ref{fig:LP400nm} shows the simulation results of the $e^-/e^+$ bunch colliding with the laser pulse for each field configuration. By comparing the total [Fig.~\ref{fig:LP400nm}\,(a)] and incoherent [Fig.~\ref{fig:LP400nm}\,(e)] spectra, frequencies at which coherent emission occurs can be identified, and then explained by examining the bunch dynamics at the laser pulse peak~[Fig.~\ref{fig:LP400nm}\,(b)]. The spectrum is characterized by a broad series of harmonics beginning at $\omega_1\approx23\,\si{\electronvolt}$ or $\lambda_1\approx55\,\si{\nano\metre}$, at which coherent emission always occurs due to the small size of the initial bunch. For `laser field only', the bunch width remains stable at $\text{FWHM}_{\text{b}}$ throughout the simulation, and one can see coherence for $\omega\lesssim 2\pi/\text{FWHM}_{\text{b}}\approx78\,\si{\electronvolt}$. For `laser \& interparticle fields', the bunch is compressed by the emitted radiation to FWHM$_\text{mb}\approx 4.4\,\si{\nano\metre}$, approximately one-quarter of its initial width. This leads to increased coherence at correspondingly high frequencies $\omega\lesssim2\pi/\text{FWHM}_{\text{mb}}\approx280\,\si{\electronvolt}$.

\begin{figure}[t]
    \centering
    \includegraphics[width=0.999\linewidth]{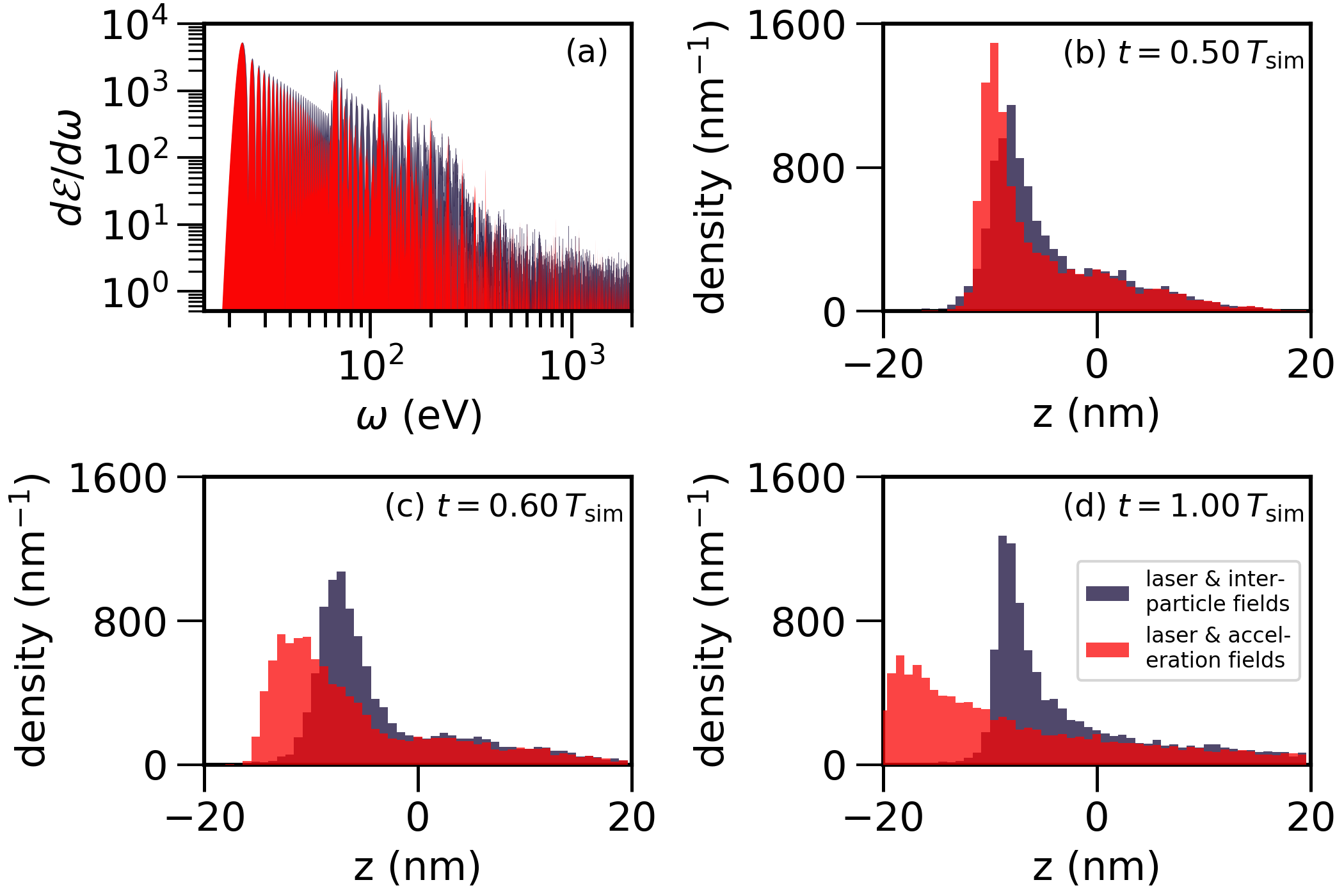}
    \caption{\textbf{Collision of $e^-/e^+$ bunch with laser pulse, with and without velocity fields.} From our point particle simulations: (a) spectrum radiated onto detector, (b) bunch at laser pulse peak, (c) bunch at intermediate time, (d) bunch at end of simulation. Legend in (d) applies to all plots, $T_{\text{sim}}$ is the simulation duration. The acceleration fields lead to the compression of the bunch, while the velocity fields between $e^-$ and $e^+$ stabilize the bunch and maintain its coherence over time.
    }
    \label{fig:LP400nm_acc}
\end{figure}

\sectionminor{Attosecond pulse train.} 
Figure~\ref{fig:LP400nm}\,(c) shows that the light emitted by this compressed bunch takes the form of an attosecond pulse train. These pulses are approximately 8\,\si{\atto\second} in duration as shown in Fig.~\ref{fig:LP400nm}\,(d), and the time interval between pulses $\Delta t_{\text{det}}\approx92\,\si{\atto\second}$ agrees with our analytical model~(see Methods section). As the radiation pules are remarkably short, and the light emitted is broadly coherent, this suggests that the spectral phase is well-controlled. Therefore, the spectral phase $\theta\equiv\theta(\omega)$ of the transversely polarized radiation emitted in forward direction is shown in Fig.~\ref{fig:LP400nm}\,(f). This can be defined from the radiation integral ${\rho(\omega) e^{i\theta(\omega)}\equiv\sum^N_{i=1}\bm{\mathcal{I}}_i(\omega, \bm{\hat{z}})\cdot\bm{\hat{x}}}$.

Notice that $\theta(\omega)$ increases linearly with frequency $\omega$ and is particularly well behaved when the interparticle fields are included and the bunch is compressed. A linear spectral phase corresponds to a temporal shift of the time-dependent signal. Therefore, a linear fit has been subtracted from $\theta$ to obtain the residual phase $\Delta\theta$ in Fig.~\ref{fig:LP400nm}\,(g). For `laser \& interparticle fields', the residual phase is essentially flat up to $\sim350\,\si{\electronvolt}$, and so we expect that the attosecond pulses are near the Fourier transform limit. Now, the chirp of an attosecond pulse is often compensated by material dispersion~\cite{kim2004}. For this reason we have plotted the group delay dispersion in Fig.~\ref{fig:LP400nm}\,(h), which is linear and near-zero below 300\,\si{\electronvolt}. This suggests that we could consider the third order derivative, however we are not aware of any material which could compensate for this term. Therefore, we assume that the attosecond pulses produced in this scheme cannot be compressed further.

\sectionminor{Role of the positrons.} 
In Figure~\ref{fig:LP400nm}, the interaction of the $e^-/e^+$ bunch with its emitted radiation leads to bunch compression and increased coherence. It is worth clarifying the role of the positrons in this process. Therefore, we have repeated this simulation without velocity fields, and the results are displayed in Fig.~\ref{fig:LP400nm_acc}. By switching off the velocity fields, we reduce the energy radiated onto the detector by approximately one-third~[Fig.~\ref{fig:LP400nm_acc}\,(a)]. To explain this, consider the longitudinal distribution of particles throughout the simulation. At the laser pulse peak~[Fig.~\ref{fig:LP400nm_acc}\,(b)] the bunch is compressed regardless of whether or not the velocity fields are included. This confirms that the acceleration fields (radiation) compress the bunch. At advanced times~[Fig.~\ref{fig:LP400nm_acc}\,(c,\,d)], the bunch remains stable and localized only when velocity fields are included. This suggests that the velocity fields provide a restoring force between oppositely charged particles, which allows for sustained coherent emission over a longer time period. Moreover, to highlight the role of the positrons, we observe that all simulations with `laser \& intraspecies fields', where the $e^-$ and $e^+$ artificially do not feel each other's fields, suffer from a Coulomb explosion which prevents bunch compression and hinders coherent emission~[see the light blue line in Fig.~\ref{fig:LP400nm}(a,b)].

\begin{figure*}[t]
    \centering
    \includegraphics[width=0.999\linewidth]{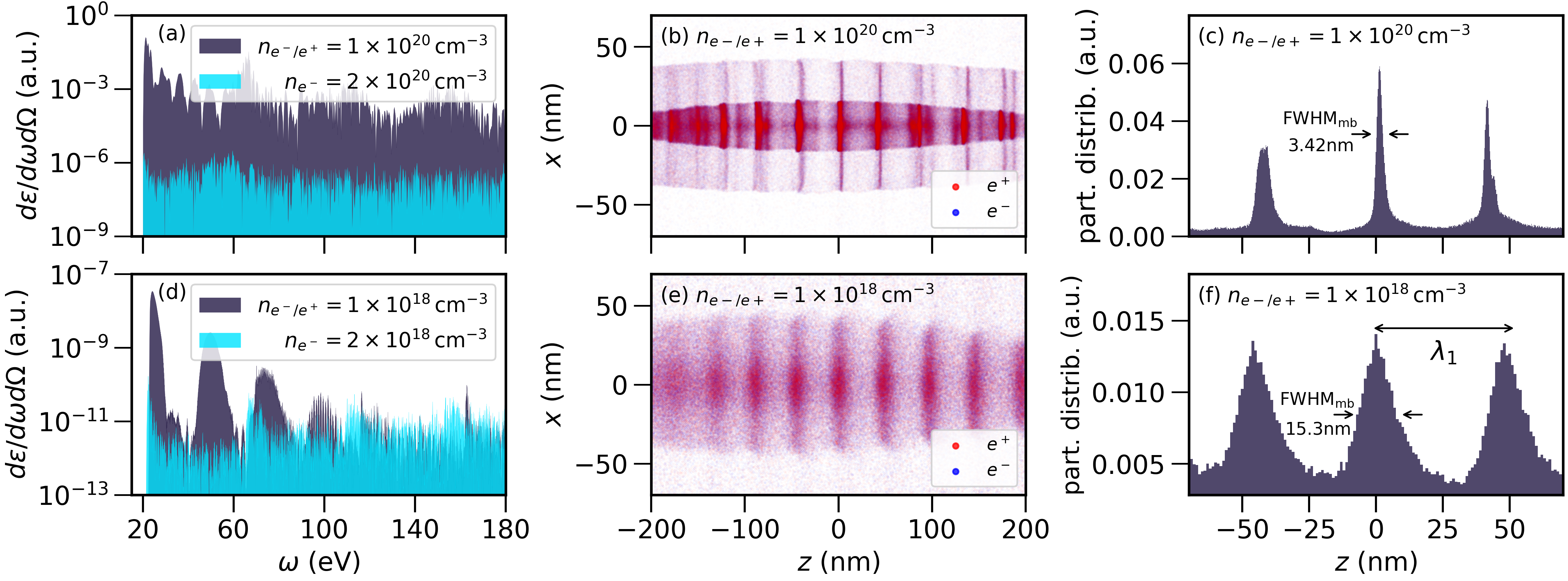}
    \caption{\textbf{3D PIC simulations of $e^-/e^+$ beam colliding with a laser pulse.} (a) radiation spectrum, (b) microbunch train, and (c) particle distribution produced from a beam of peak density $10^{20}\,\text{particle}/\si{\centi\metre^3}$ colliding with a FWHM$_L=100\,\si{\femto\second}$ laser pulse. This is repeated in (d,\,e,\,f) with a lower density beam $10^{18}\,\text{particle}/\si{\centi\metre^3}$ colliding with a longer pulse FWHM$_L=400\,\si{\femto\second}$. The spectra are observed along the $+z$ direction and the microbunches are pictured shortly after the laser pulse peak. Coherent emission occurs due to the formation of nanoscale microbunches separated by approximately $\lambda_1\approx55\,\si{\nano\metre}$. When the simulations are repeated with an electron beam, no microbunching or coherent emission occurs. Note that ``a.u.'' denotes arbitrary units.
    }
    \label{fig:pic_results}
\end{figure*}

\sectionminor{Particle-in-cell simulations.}
The simulations performed above with our first-principles point particle code highlight how the stability of $e^-/e^+$ microbunches allows them to emit broadband coherent light in the form of attosecond pulses. By performing fully three-dimensional PIC simulations with the Smilei code~(see Methods section), we show how a train of similar microbunches can be created when an $e^-/e^+$ beam interacts with its own radiation upon colliding with a laser pulse. 

Consider a Gaussian $e^-/e^+$ beam of width FWHM$_\perp\approx 0.16\,\si{\micro\metre}$ and length FWHM$_\parallel\approx 1\,\si{\micro\metre}$. As before, the beam has an average kinetic energy of $2.0\,\si{\mega\electronvolt}$, divergence $\sigma_\vartheta=1\,\si{\milli\radian}$, and kinetic energy spread $\sigma_{\text{KE}}=0.1\,\%$, and collides head-on with a plane wave pulse of amplitude $a_0=5$ and central wavelength $\lambda_0=400\,\si{\nano\metre}$. In our first simulation, the beam has a peak density of $n_{e^-/e^+}=10^{20}\,\text{particles}/\si{\centi\metre^3}$ for each species and collides with a pulse of duration FWHM$_L=100\,\si{\femto\second}$. In our second simulation, the beam has a lower density $n_{e^-/e^+}=10^{18}\,\text{particles}/\si{\centi\metre^3}$ and collides with a longer pulse FWHM$_L=400\,\si{\femto\second}$. Both simulations are repeated with a beam containing no positrons and twice the density of electrons, to compare beams with the same total number of particles, demonstrating the advantage of including positrons.

Electron beams with peak densities approaching $n_{e^-}=10^{19}\,\text{particles}/\si{\centi\metre^3}$ have been produced via laser-wakefield acceleration~\cite{liuPoP23}, while electron-positron beams with peak density $n_{e^-/e^+}=10^{16}\,\text{particles}/\si{\centi\metre^3}$ and divergence 10--20\,\si{\milli\radian} have been produced from a compact laser-driven setup~\cite{sarri2015}. It is well known from FEL theory that the gain length varies inversely with the beam density~\cite{schmueser2009, steiniger2014}. Consequently, a longer pulse duration is needed to develop a microbunch train when employing a lower density of particles. Therefore, this scheme could potentially operate at densities below $n_{e^-/e^+}=10^{18}\,\text{particles}/\si{\centi\metre^3}$, if a longer laser pulse was employed. However, simulating longer pulses significantly increases the computational cost of PIC simulations. 

\sectionminor{Train of electron-positron microbunches.}
Figure~\ref{fig:pic_results} shows the results of the beam and laser pulse collision. In Fig.~\ref{fig:pic_results}\,(a), we can see that the dense beam with $n_{e^-/e^+}=10^{20}\,\text{particles}/\si{\centi\metre^3}$ is capable of producing broadband coherent light well into the XUV region, similar to the spectrum emitted by the $e^-/e^+$ bunch in Fig.\,\ref{fig:LP400nm}\,(a). The additional coherence caused by the presence of the positrons is explained by the formation of a dense microbunch train in Fig.~\ref{fig:pic_results}\,(b) shortly after the laser pulse peak. The microbunches are separated approximately by the first harmonic $\lambda_1$ in Fig.~\ref{fig:pic_results}\,(c), and have a marginally smaller width $\text{FWHM}_{\text{mb}}=3.42\,\si{\nano\metre}$ and higher coherence than the compressed bunch studied in Fig.~\ref{fig:LP400nm}\,(b). We conclude that the individual microbunches are capable of emitting an attosecond pulse train as shown in Fig.~\ref{fig:LP400nm}\,(c,\,d). At the high particle density considered here, the microbunches form rapidly, emit coherently, and then expand, all over a distance of less than $50\,\si{\micro\metre}$. When this simulation was repeated with an electron beam of the same density in Fig.~\ref{fig:pic_results}\,(a), no microbunching or coherent emission occurred; this is comparable to the simulations with `laser \& intraspecies fields' shown earlier.

As it is challenging to produce a high-density $e^-/e^+$ beam in the laboratory, we have repeated this simulation at a far lower density $n_{e^-/e^+}=10^{18}\,\text{particles}/\si{\centi\metre^3}$ (see Discussion section). 
In this case, the spectrum in Fig.~\ref{fig:pic_results}\,(d) is dominated by coherent emission around the first few harmonics. The prominence of the individual harmonics can be explained by the longer and consequently more monochromatic laser pulse employed in this simulation. Despite lowering the density by two orders of magnitude, we still observe an impressive train of nanoscale microbunches in Fig.~\ref{fig:pic_results}\,(e) of width FWHM$_{\text{mb}}=15.3\,\si{\nano\metre}$ or equivalently 51\,\si{\atto\second}. The microbunch separation occurs approximately at the first harmonic $\lambda_1$, and the microbunch width effectively controls the range of frequencies at which coherent emission occurs $\omega\lesssim 2\pi/\text{FWHM}_{\text{mb}}\approx80\,\si{\electronvolt}$. Once again, we have repeated this simulation with an electron beam of the same density in Fig.~\ref{fig:pic_results}\,(d). Despite considering a much lower density where Coulomb repulsion should play a less significant role, we still observe no microbunching or coherent emission unless the positrons are present. This underlines the need for a neutral particle beam to create a functioning OFEL.

An additional simulation was performed to understand the effect of the initial energy spread on the results in Fig.~\ref{fig:pic_results}, as it is one of the key parameters describing beam quality and poses a major challenge in the beam production chain. In particular, we found that increasing the kinetic energy spread to $\sigma_{\text{KE}}=0.6\,\%$ had little impact on the coherence at the first harmonic, as shown in Fig.~\ref{fig:pic_results}\,(a), which dominates the system. However, it did reduce the amplitude of the higher harmonics by roughly one order of magnitude. This behavior corresponds to the larger FWHM$_{\text{mb}}$, which acts as a low-pass filter on the coherence at higher harmonics. This effect can also be thought of as an increase in the FEL gain length.

\sectionmajor{Discussion}
\smallskip
\\
\noindent
In this paper, we presented simulations of the collision between a neutral, relativistic $e^-/e^+$ bunch and a laser pulse. This bunch is compressed by self-interaction with its radiation and emits coherent light up to about $280\,\si{\electronvolt}$ in the form of $8\,\si{\atto\second}$ pulses. Also, we have shown how a train of microbunches with similar width and coherence could be produced spontaneously in a dense, micron-scale $e^-/e^+$ beam. The spectral and temporal properties of the emitted radiation compare favorably with state-of-the-art experimental results from high harmonic generation and attosecond-FELs, which are capable of generating pulses of about $40\,\si{\atto\second}$~\cite{gaumnitz2017} and $200\,\si{\atto\second}$~\cite{duris2020} duration, respectively. It is clear that the inherent symmetry of an $e^-/e^+$ system, with particles of equal mass and opposite charge, allows for the realization of a compact source of coherent light.

The impressive properties of the emitted light must be weighed against the challenge of producing an $e^-/e^+$ beam of density $n_{e^-/e^+}=10^{20}\,\text{particles}/\si{\centi\metre^3}$. To address this, we repeated our simulation using a much lower-density beam, $n_{e^-/e^+}=10^{18}\,\text{particles}/\si{\centi\metre^3}$, colliding with a longer laser pulse. Nevertheless, this collision produced an impressive train of nanoscale microbunches, of 51\,\si{\atto\second} duration, which emit coherent XUV light. The challenge here, as for conventional FELs~\cite{pellegriniRMP16}, lies in producing a high-quality particle beam needed to realize this scheme. For the particle densities considered here, we have demonstrated that microbunching with electrons alone in an OFEL is remarkably difficult due to Coulomb repulsion and the short interaction time with the laser pulse; a neutral beam will be necessary to observe microbunching over microscopic distances. The primary difficulty will be in producing a dense $e^-/e^+$ beam with a narrow divergence and energy spread. 

We emphasize that high-density ($n_{e^-/e^+}=10^{16}\,\text{particles}/\si{\centi\metre^3}$) and low-divergence ($\sim10$~mrad) neutral electron–positron beams with multi-MeV energies ($\gamma_0 \approx 15$) have been produced in the laboratory~\cite{sarri2015}, and electron beams with peak densities approaching $n_{e^-}=10^{19}\,\text{particles}/\si{\centi\metre^3}$ have been achieved via laser-wakefield acceleration~\cite{liuPoP23}. Moreover, ongoing advances in laser technology have enabled the production of narrow-bandwidth, low-emittance positron beams from laser-wakefield accelerators~\cite{streeter2024}; the active energy compression of laser-plasma electron beams to below the permille-level energy spread~\cite{winkler2025}; the experimental demonstration of free-electron lasing with a compact beam-driven plasma wakefield accelerator~\cite{pompili2022}; and the modulation of electron energy profiles to shape ultrashort (femtosecond-duration), ultra-high-current (0.1~MA) electron beams in particle accelerators~\cite{emmaPRL25}. These developments suggest that, while the extraordinary results obtained with the $e^-/e^+$ beam of density $n_{e^-/e^+}=10^{20}\,\text{particles}/\si{\centi\metre^3}$ represent a long-term goal, those achieved with $n_{e^-/e^+}=10^{18}\,\text{particles}/\si{\centi\metre^3}$ could be realized using current technology. These results, therefore, constitute a proof of principle of a tabletop source of temporally coherent broadband XUV radiation, with broad applications in physics, chemistry, and biology, as well as in industry.

\sectionmajor{Methods}

\sectionminor{Point particle code.} Equations~\eqref{eq:Fret}--\eqref{eq:energy} are solved numerically as follows: our code is initialized by assuming the particles propagate ballistically before colliding with the laser pulse (a sensible assumption for a neutral bunch). The trajectories are stored at discrete time steps in the memory, and the fields in Eqs.~\eqref{eq:Fret}--\eqref{eq:Facc} are evaluated at the retarded time(s) by interpolation. Then, the reduced Landau-Lifshitz equation~\eqref{eq:LLeq} is integrated with a second-order leapfrog scheme~\cite{tamburini2010}. Finally, the radiation spectrum~\eqref{eq:energy} is found via a fast Fourier transform. This approach allows the radiation properties and bunch compression to be studied in detail from first principles~[see Fig.~\ref{fig:LP400nm}]. In particular, the ability of our code to distinguish between the acceleration and velocity fields allows us to explain the stability of the $e^-/e^+$ system~[Fig.~\ref{fig:LP400nm_acc}]. However, the number of particles $N$ and the spatial size of the bunch that can be simulated is limited by the need to keep the entire trajectories of all particles in the memory. The computational cost $O(N^2)$ increases rapidly with $N$, and hence we are limited to $N\lesssim10^4$ given the resources available. Further information and tests of the code can be found in Ref.~\cite{quin2023_phd}. We also acknowledge that the microbunching and spectrum obtained with our point particle code were reproduced~\cite{almeidal2023} with the PIC code OSIRIS~\cite{fonesca2002}.

\sectionminor{Particle-in-cell simulation parameters.}
The PIC simulations are performed in 3D Cartesian geometry using the Smilei code~\cite{derouillat2018}. Electromagnetic field initialization is carried out with a relativistic Poisson solver, the standard Yee scheme is used as a Maxwell solver, and the trajectories are found by solving the Landau-Lifshitz equation with an applied Boris pusher. We employ open (absorbing) and periodic boundary conditions for the particles in the longitudinal and transverse directions, respectively, and absorbing boundary conditions for the fields.

The corresponding cell sizes are $\Delta z=4\,\si{\nm}$ and $\Delta x=\Delta y=10\,\si{\nm}$, while the imposed CFL\,-\,condition is 0.95. By considering Nyquist's theorem, we aimed to provide sufficient longitudinal sampling (4 per optical cycle) up to the 3rd harmonic while providing moderate resolution in transverse direction to avoid numerical heating. The simulation box length $L_z\approx2\,\text{FWHM}_L$ and width $L_x=L_y=1.28\,\si{\um}$ are determined by the need to fully resolve the collision and avoid interference from radiation which is not fully absorbed at the boundaries. The particle beam is represented by the $10^7$ macro-particles per species with an uniform weight distribution and a Gaussian distribution in the phase space. To retrieve the far-field power spectrum, we have implemented the RaDiO~\cite{pardal2023radio} algorithm in the Smilei framework in a way that keeps performance optimal and does not disrupt the parallelism of the code.

\sectionminor{Plane wave approximation.} 
In our point particle simulations, we model the laser pulse as a linearly-polarized plane wave with vector potential 
%$
\begin{align}
	\frac{|e|}{m}\bm{A}_{\text{ext}}(\varphi) &= a(\varphi)\,\sin(\varphi)\,\hat{\bm{x}},
	\\
	a(\varphi) &= a_0\cos^2(\varphi/\Delta).
\end{align}
The four-potential $A^\mu_{\text{ext}}(\varphi)=(0, \bm{A}_{\text{ext}}(\varphi))$ is chosen to satisfy the Lorenz gauge $(\partial A_{\text{ext}}(\varphi))=0$. The pulse propagates along $-z$ with wave phase $\varphi=\omega_0z_+$, where ${z_\pm=t\pm z}$ are the light-cone coordinates. The envelope satisfies $a(\varphi_0)=0$ at the initial phase $\varphi_0=-\pi\Delta/2$, and the envelope domain is ${\varphi\in[-\pi\Delta/2, +\pi\Delta/2]}$. 

In the regime of interest here, where $a_0\approx \gamma_0$, the plane wave approximation is valid when the laser waist is much larger than the transverse bunch width. For example, the results as plotted in Fig.~\ref{fig:LP400nm} were essentially unchanged when this simulation was repeated with a focused laser pulse of waist radius $w_0=4\,\si{\micro\metre}$. For the focused laser field employed see Ref.~\cite{salamin2002}, where we have considered terms up to third order in the diffraction angle.

\sectionminor{Trajectory in a plane wave.}
Consider a particle colliding head-on with the external plane wave pulse. Neglecting radiation reaction, the four-velocity can be written in terms of the light-cone coordinates $u_\pm \equiv \gamma\pm u_z$~\cite{landaulifshitz_vol2,dipiazza2012_review}
\begin{align}
	\bm{u}_{\perp}(\varphi) &= -\frac{e}{m} \bm{A}_{\text{ext}}(\varphi),
	\label{eq:u_perp}
	\\
	u_-(\varphi) &= \frac{1}{u_{0,+}}\left[1+\bm{u}^2_\perp(\varphi)\right],
	\label{eq:u_minus}
	\\
	u_+(\varphi) &= u_{0,+}.
	\label{eq:u_plus}
\end{align}
In our simulations, the trajectories begin to differ from Eqs.\,\eqref{eq:u_perp}--\eqref{eq:u_plus} when the cumulative effect of interparticle fields, i.e. microbunching, becomes important.

\sectionminor{Model of microbunching.}
Here we demonstrate that resonance, energy transfer and microbunching can occur at the harmonics of reflected light in a plane wave. The method employed is similar to explanations of microbunching in a low gain FEL~\cite{schmueser2009}. Since the aim is to show at which frequencies most of the energy is transferred from the electrons to the radiation, we start from the evolution of the Lorentz factor for a single particle
\begin{equation}
	\frac{d\gamma}{d\varphi} = \frac{e\bm{E}(\varphi)\cdot\bm{u}(\varphi)}{m\omega_0u_{0,+}},
	\label{eq:lorentz_gamma}
\end{equation}
where we have used $d\varphi/d\tau=\omega_0 u_{0,+}$ with $\tau$ being the proper time. As the $e^-/e^+$ bunch is quasi-neutral throughout the simulation we can neglect Coulomb fields here. Then, the total field depends on the external laser field and a radiation field ${\bm{E}(\varphi) = \bm{E}_{\text{ext}}(\varphi) + \bm{E}_{\text{rad}}(\varphi)}$. We assume the radiation field can be approximated as an arbitrary series of plane waves propagating along $+z$ with transverse polarization
\begin{equation}
	\bm{E}_{\text{rad}}(\varphi) = \sum_{l} \bm{E}_{l,\perp} \sin\phi_l(\varphi),
\end{equation}
where $\phi_l(\varphi)=\omega_l z_-(\varphi) + \psi_l$ is the radiation wave phase, $\omega_l$ is the frequency, and $\psi_l$ is an arbitrary constant phase. In the monochromatic approximation $\Delta\gg1$ we have $a(\varphi)\approx a_0$. For example, in our simulation with $\lambda_0=400\,\si{\nano\metre}$ and FWHM$_L\approx 26.7\,\si{\femto\second}$ we have $\Delta\approx110$. 
One can then derive an expression for the position ${z_-(\varphi)=z_{0,-}+\int^\varphi_{\varphi_0}[u_-(\varphi')/(\omega_0 u_{0,+})]d\varphi'}$, to obtain
\begin{align}
    z_-(\varphi) - z_{0,-} = &\frac{1}{\omega_0u^2_{0,+}} \Bigg[\Bigg. \left(1+\frac{a_0^2}{2}\right) \left(\varphi -\varphi_0 \right) \nonumber 
    \\ 
    & - \frac{a^2_0}{4} \left[ \sin(2\varphi) - \sin(2\varphi_0) \right]\Bigg.\Bigg].
    \label{eq:x_minus}
\end{align}
By comparing the particles' trajectories in our simulations to the exact solutions in a plane wave [Eqs.\,\eqref{eq:u_perp}--\eqref{eq:u_plus}], we know that the external field dominates the system $|\bm{E}_{\text{rad}}(\varphi)|\ll |\bm{E}_{\text{ext}}(\varphi)|$, and therefore the effect of the radiation field on the trajectory can be treated perturbatively. Under the assumption that the particles' trajectory is determined by the plane wave, on average, there is no energy exchanged between the particle and external plane-wave field $\langle\bm{E}_{\text{ext}}(\varphi)\cdot\bm{u}_\perp\rangle=0$. Here, the cycle-average is defined as ${\langle f(\varphi)\rangle = \frac{1}{2\pi}\int^{\varphi+\pi}_{\varphi-\pi} f(\varphi') d\varphi'}$ for a generic function $f(\varphi)$. The average energy exchanged between the particle and radiation field is then
\begin{equation}
	\left\langle\frac{d\gamma}{d\varphi}\right\rangle = \sum_l \frac{a_0|e|E_{l,x}}{2m\omega_0 u_{0,+}} \operatorname{Re}\left[ \big\langle e^{i\eta_+^l(\varphi)} \big\rangle - \big\langle e^{i\eta_-^l(\varphi)} \big\rangle \right],
    \label{eq:average}
\end{equation}
where $\eta^l_\pm(\varphi)=\phi_l(\varphi)\pm\varphi$ and $E_{l,x}=\bm{E}_{l,\perp}\cdot\bm{\hat{x}}$. One can solve these integrals using Bessel functions
%of the first kind
%
\begin{equation}
    \operatorname{Re}\,\big\langle e^{i\eta^\pm_l(\varphi)} \big\rangle = \sum^{\infty}_{n=-\infty} J_n(\rho_l) \frac{\sin(\pi \Theta^\pm_{l,n})}{\pi \Theta^\pm_{l,n}} \, \cos\left(\Theta^{l,n}_\pm\varphi + \psi'_l\right),
\end{equation}
where $\rho_l = \omega_l a_0^2/4\omega_0 u^2_{0,+}$, and we have used the generating function~\cite{abramowitz1968}
\begin{equation}
    e^{-i\rho_l \sin(2\varphi)} = \sum^{\infty}_{n=-\infty} J_n(\rho_l)\,e^{-2 i n \varphi}.
\end{equation}
Note that the coefficient of $\varphi$ in the phase of each harmonic is given by 
\begin{equation}
	\Theta^{l,n}_\pm = \frac{\omega_l}{u^2_{0,+}\omega_0} \left(1+\frac{a^2_0}{2}\right) - 2n \pm 1,
\end{equation}
and $\psi'_l$ is the modified constant phase, defined as
\begin{equation}
    \psi'_l = \psi_l + \omega_l z_{0,-} -
    \frac{\omega_l}{\omega_0 u^2_{0,+}} \left[\left(1+\frac{a_0^2}{2}\right) \varphi_0 - \frac{a_0^2 \sin(2\varphi_0)}{4} \right].
\end{equation}
For each frequency $\omega_l$, the dominant contribution from the sum over $n$ occurs at $\Theta^{l,n}_\pm=0$. If averaged over an infinite number of cycles, instead of a single cycle as above, a delta function would collapse the sum over $n$ leaving only terms where $n=l$. We conclude that resonance occurs at $\Theta^{l,n}_+=0$, such that (taking $n=l$)
\begin{equation}
	\frac{\omega_l}{4\gamma^2_0\omega_0} = \frac{2l-1}{1+\frac{1}{2}a^2_0},
	\label{eq:harmonics}
\end{equation} 
and for a relativistic particle $u^2_{0,+}\approx4\gamma_0^2$. These are exactly the harmonics emitted in a monochromatic plane wave~\cite{sarachik1970, salamin1996}. In practice, $\lambda_1=2\pi/\omega_1$ is the longest emitted harmonic, and emission at this wavelength scales coherently with $N^2$ such that it dominates the system and any subsequent microbunching.

We note a limitation of this model: our assumption that $\bm{E}_{l,\perp}$ is constant. From our simulations in Fig.~\ref{fig:LP400nm}\,(c) with `laser \& interparticle fields', the amplitude of the radiation field actually increases over time, as opposed to remaining constant. Therefore, this model cannot predict the intensity of radiation emitted quantitatively, but does suffice to predict the resonant wavelength at which microbunching occurs.

\sectionminor{Time interval between radiation pulses.}
The light-cone coordinate $z_-(\varphi)$ in Eq.\,\eqref{eq:x_minus} can be interpreted as the time measured by the detector. For one oscillation of the plane wave, the particle's radiation cone will sweep across the detector twice. Therefore, the time period between radiation pulses is the change in $z_-(\varphi)$ over half a cycle
\begin{equation}
    \Delta t_{\text{det}} = z_-(\varphi+\pi) - z_-(\varphi) \approx \frac{T_0}{8\gamma_0^2} \left(1+\frac{a^2_0}{2}\right).
    \label{eq:dtdet}
\end{equation}
Here $T_0=2\pi/\omega_0$ is the plane wave period and we have assumed the particle is relativistic. For example, with wavelength $\lambda_0=400\,\si{\nano\metre}$ one can expect radiation pulses separated by a time interval $\Delta t_{\text{det}}\approx92\,\si{\atto\second}$ in excellent agreement with Fig.~\ref{fig:LP400nm}\,(d).%\\

\sectionmajor{Data availability}
\smallskip

\noindent
The data and Python scripts required to reproduce the figures presented here are available on Zenodo~\cite{figuredata}.

\sectionmajor{Code availability}
\smallskip

\noindent
The key results of this paper have been produced with the publicly available particle-in-cell code Smilei. The working principles of our point-particle code have been outlined in this paper, and we have included references where one can find further details and tests of the code. In future, we aim to make this code available for public use; before then, it can be made available upon reasonable request to M.J.Q.%\\

\sectionmajor{References}
\bibliography{bibliography}

\begin{thebibliography}{59}%
\makeatletter
\providecommand \@ifxundefined [1]{%
 \@ifx{#1\undefined}
}%
\providecommand \@ifnum [1]{%
 \ifnum #1\expandafter \@firstoftwo
 \else \expandafter \@secondoftwo
 \fi
}%
\providecommand \@ifx [1]{%
 \ifx #1\expandafter \@firstoftwo
 \else \expandafter \@secondoftwo
 \fi
}%
\providecommand \natexlab [1]{#1}%
\providecommand \enquote  [1]{``#1''}%
\providecommand \bibnamefont  [1]{#1}%
\providecommand \bibfnamefont [1]{#1}%
\providecommand \citenamefont [1]{#1}%
\providecommand \href@noop [0]{\@secondoftwo}%
\providecommand \href [0]{\begingroup \@sanitize@url \@href}%
\providecommand \@href[1]{\@@startlink{#1}\@@href}%
\providecommand \@@href[1]{\endgroup#1\@@endlink}%
\providecommand \@sanitize@url [0]{\catcode `\\12\catcode `\$12\catcode
  `\&12\catcode `\#12\catcode `\^12\catcode `\_12\catcode `\%12\relax}%
\providecommand \@@startlink[1]{}%
\providecommand \@@endlink[0]{}%
\providecommand \url  [0]{\begingroup\@sanitize@url \@url }%
\providecommand \@url [1]{\endgroup\@href {#1}{\urlprefix }}%
\providecommand \urlprefix  [0]{URL }%
\providecommand \Eprint [0]{\href }%
\providecommand \doibase [0]{https://doi.org/}%
\providecommand \selectlanguage [0]{\@gobble}%
\providecommand \bibinfo  [0]{\@secondoftwo}%
\providecommand \bibfield  [0]{\@secondoftwo}%
\providecommand \translation [1]{[#1]}%
\providecommand \BibitemOpen [0]{}%
\providecommand \bibitemStop [0]{}%
\providecommand \bibitemNoStop [0]{.\EOS\space}%
\providecommand \EOS [0]{\spacefactor3000\relax}%
\providecommand \BibitemShut  [1]{\csname bibitem#1\endcsname}%
\let\auto@bib@innerbib\@empty
%</preamble>
\bibitem [{\citenamefont {Krausz}\ and\ \citenamefont
  {Ivanov}(2009)}]{krausz2009}%
  \BibitemOpen
  \bibfield  {author} {\bibinfo {author} {\bibfnamefont {F.}~\bibnamefont
  {Krausz}}\ and\ \bibinfo {author} {\bibfnamefont {M.}~\bibnamefont
  {Ivanov}},\ }\bibfield  {title} {\bibinfo {title} {Attosecond physics},\
  }\href {https://doi.org/10.1103/RevModPhys.81.163} {\bibfield  {journal}
  {\bibinfo  {journal} {Rev. Mod. Phys.}\ }\textbf {\bibinfo {volume} {81}},\
  \bibinfo {pages} {163} (\bibinfo {year} {2009})}\BibitemShut {NoStop}%
\bibitem [{\citenamefont {Calegari}\ \emph {et~al.}(2016)\citenamefont
  {Calegari}, \citenamefont {Sansone}, \citenamefont {Stagira}, \citenamefont
  {Vozzi},\ and\ \citenamefont {Nisoli}}]{calegari2016}%
  \BibitemOpen
  \bibfield  {author} {\bibinfo {author} {\bibfnamefont {F.}~\bibnamefont
  {Calegari}}, \bibinfo {author} {\bibfnamefont {G.}~\bibnamefont {Sansone}},
  \bibinfo {author} {\bibfnamefont {S.}~\bibnamefont {Stagira}}, \bibinfo
  {author} {\bibfnamefont {C.}~\bibnamefont {Vozzi}},\ and\ \bibinfo {author}
  {\bibfnamefont {M.}~\bibnamefont {Nisoli}},\ }\bibfield  {title} {\bibinfo
  {title} {Advances in attosecond science},\ }\href
  {https://doi.org/10.1088/0953-4075/49/6/062001} {\bibfield  {journal}
  {\bibinfo  {journal} {J. Phys. B.: At. Mol. Opt. Phys.}\ }\textbf {\bibinfo
  {volume} {49}},\ \bibinfo {pages} {062001} (\bibinfo {year}
  {2016})}\BibitemShut {NoStop}%
\bibitem [{\citenamefont {L{\'e}pine}\ \emph {et~al.}(2014)\citenamefont
  {L{\'e}pine}, \citenamefont {Ivanov},\ and\ \citenamefont
  {Vrakking}}]{lepine2014}%
  \BibitemOpen
  \bibfield  {author} {\bibinfo {author} {\bibfnamefont {F.}~\bibnamefont
  {L{\'e}pine}}, \bibinfo {author} {\bibfnamefont {M.~Y.}\ \bibnamefont
  {Ivanov}},\ and\ \bibinfo {author} {\bibfnamefont {M.~J.~J.}\ \bibnamefont
  {Vrakking}},\ }\bibfield  {title} {\bibinfo {title} {Attosecond molecular
  dynamics: fact or fiction?},\ }\href
  {https://doi.org/10.1038/nphoton.2014.25} {\bibfield  {journal} {\bibinfo
  {journal} {Nat. Photonics}\ }\textbf {\bibinfo {volume} {8}},\ \bibinfo
  {pages} {195} (\bibinfo {year} {2014})}\BibitemShut {NoStop}%
\bibitem [{\citenamefont {Nisoli}\ \emph {et~al.}(2017)\citenamefont {Nisoli},
  \citenamefont {Decleva}, \citenamefont {Calegari}, \citenamefont {Palacios},\
  and\ \citenamefont {Mart{\'i}n}}]{nisoli2017}%
  \BibitemOpen
  \bibfield  {author} {\bibinfo {author} {\bibfnamefont {M.}~\bibnamefont
  {Nisoli}}, \bibinfo {author} {\bibfnamefont {P.}~\bibnamefont {Decleva}},
  \bibinfo {author} {\bibfnamefont {F.}~\bibnamefont {Calegari}}, \bibinfo
  {author} {\bibfnamefont {A.}~\bibnamefont {Palacios}},\ and\ \bibinfo
  {author} {\bibfnamefont {F.}~\bibnamefont {Mart{\'i}n}},\ }\bibfield  {title}
  {\bibinfo {title} {Attosecond electron dynamics in molecules},\ }\href
  {https://doi.org/10.1021/acs.chemrev.6b00453} {\bibfield  {journal} {\bibinfo
   {journal} {Chem. Rev.}\ }\textbf {\bibinfo {volume} {117}},\ \bibinfo
  {pages} {10760} (\bibinfo {year} {2017})}\BibitemShut {NoStop}%
\bibitem [{\citenamefont {Ciappina}\ \emph {et~al.}(2017)\citenamefont
  {Ciappina}, \citenamefont {P{\'e}rez-Hern{\'a}ndez}, \citenamefont
  {Landsman}, \citenamefont {Okell}, \citenamefont {Zherebtsov}, \citenamefont
  {F{\"o}rg}, \citenamefont {Sch{\"o}tz}, \citenamefont {Seiffert},
  \citenamefont {Fennel}, \citenamefont {Shaaran}, \citenamefont {Zimmermann},
  \citenamefont {Chac{\'o}n}, \citenamefont {Guichard}, \citenamefont
  {Za{\"i}r}, \citenamefont {Tisch}, \citenamefont {Marangos}, \citenamefont
  {Witting}, \citenamefont {Braun}, \citenamefont {Maier}, \citenamefont
  {Roso}, \citenamefont {Kr{\"u}ger}, \citenamefont {Hommelhoff}, \citenamefont
  {Kling}, \citenamefont {Krausz},\ and\ \citenamefont
  {Lewenstein}}]{ciappina2017}%
  \BibitemOpen
  \bibfield  {author} {\bibinfo {author} {\bibfnamefont {M.~F.}\ \bibnamefont
  {Ciappina}}, \bibinfo {author} {\bibfnamefont {J.~A.}\ \bibnamefont
  {P{\'e}rez-Hern{\'a}ndez}}, \bibinfo {author} {\bibfnamefont {A.~S.}\
  \bibnamefont {Landsman}}, \bibinfo {author} {\bibfnamefont {W.~A.}\
  \bibnamefont {Okell}}, \bibinfo {author} {\bibfnamefont {S.}~\bibnamefont
  {Zherebtsov}}, \bibinfo {author} {\bibfnamefont {B.}~\bibnamefont
  {F{\"o}rg}}, \bibinfo {author} {\bibfnamefont {J.}~\bibnamefont
  {Sch{\"o}tz}}, \bibinfo {author} {\bibfnamefont {L.}~\bibnamefont
  {Seiffert}}, \bibinfo {author} {\bibfnamefont {T.}~\bibnamefont {Fennel}},
  \bibinfo {author} {\bibfnamefont {T.}~\bibnamefont {Shaaran}}, \bibinfo
  {author} {\bibfnamefont {T.}~\bibnamefont {Zimmermann}}, \bibinfo {author}
  {\bibfnamefont {A.}~\bibnamefont {Chac{\'o}n}}, \bibinfo {author}
  {\bibfnamefont {R.}~\bibnamefont {Guichard}}, \bibinfo {author}
  {\bibfnamefont {A.}~\bibnamefont {Za{\"i}r}}, \bibinfo {author}
  {\bibfnamefont {J.~W.~G.}\ \bibnamefont {Tisch}}, \bibinfo {author}
  {\bibfnamefont {J.~P.}\ \bibnamefont {Marangos}}, \bibinfo {author}
  {\bibfnamefont {T.}~\bibnamefont {Witting}}, \bibinfo {author} {\bibfnamefont
  {A.}~\bibnamefont {Braun}}, \bibinfo {author} {\bibfnamefont {S.~A.}\
  \bibnamefont {Maier}}, \bibinfo {author} {\bibfnamefont {L.}~\bibnamefont
  {Roso}}, \bibinfo {author} {\bibfnamefont {M.}~\bibnamefont {Kr{\"u}ger}},
  \bibinfo {author} {\bibfnamefont {P.}~\bibnamefont {Hommelhoff}}, \bibinfo
  {author} {\bibfnamefont {M.~F.}\ \bibnamefont {Kling}}, \bibinfo {author}
  {\bibfnamefont {F.}~\bibnamefont {Krausz}},\ and\ \bibinfo {author}
  {\bibfnamefont {M.}~\bibnamefont {Lewenstein}},\ }\bibfield  {title}
  {\bibinfo {title} {Attosecond physics at the nanoscale},\ }\href
  {https://doi.org/10.1088/1361-6633/aa574e} {\bibfield  {journal} {\bibinfo
  {journal} {Rep. Prog. Phys.}\ }\textbf {\bibinfo {volume} {80}},\ \bibinfo
  {pages} {054401} (\bibinfo {year} {2017})}\BibitemShut {NoStop}%
\bibitem [{\citenamefont {Calegari}\ and\ \citenamefont
  {Martin}(2023)}]{calegari2023}%
  \BibitemOpen
  \bibfield  {author} {\bibinfo {author} {\bibfnamefont {F.}~\bibnamefont
  {Calegari}}\ and\ \bibinfo {author} {\bibfnamefont {F.}~\bibnamefont
  {Martin}},\ }\bibfield  {title} {\bibinfo {title} {Open questions in
  attochemistry},\ }\href {https://doi.org/10.1038/s42004-023-00989-0}
  {\bibfield  {journal} {\bibinfo  {journal} {Commun. Chem.}\ }\textbf
  {\bibinfo {volume} {6}},\ \bibinfo {pages} {184} (\bibinfo {year}
  {2023})}\BibitemShut {NoStop}%
\bibitem [{\citenamefont {Hentschel}\ \emph {et~al.}(2001)\citenamefont
  {Hentschel}, \citenamefont {Kienberger}, \citenamefont {Spielmann},
  \citenamefont {Reider}, \citenamefont {Milosevic}, \citenamefont {Brabec},
  \citenamefont {Corkum}, \citenamefont {Heinzmann}, \citenamefont {Drescher},\
  and\ \citenamefont {Krausz}}]{hentschel2001}%
  \BibitemOpen
  \bibfield  {author} {\bibinfo {author} {\bibfnamefont {M.}~\bibnamefont
  {Hentschel}}, \bibinfo {author} {\bibfnamefont {R.}~\bibnamefont
  {Kienberger}}, \bibinfo {author} {\bibfnamefont {C.}~\bibnamefont
  {Spielmann}}, \bibinfo {author} {\bibfnamefont {G.~A.}\ \bibnamefont
  {Reider}}, \bibinfo {author} {\bibfnamefont {N.}~\bibnamefont {Milosevic}},
  \bibinfo {author} {\bibfnamefont {T.}~\bibnamefont {Brabec}}, \bibinfo
  {author} {\bibfnamefont {P.}~\bibnamefont {Corkum}}, \bibinfo {author}
  {\bibfnamefont {U.}~\bibnamefont {Heinzmann}}, \bibinfo {author}
  {\bibfnamefont {M.}~\bibnamefont {Drescher}},\ and\ \bibinfo {author}
  {\bibfnamefont {F.}~\bibnamefont {Krausz}},\ }\bibfield  {title} {\bibinfo
  {title} {Attosecond metrology},\ }\href {https://doi.org/10.1038/35107000}
  {\bibfield  {journal} {\bibinfo  {journal} {Nature}\ }\textbf {\bibinfo
  {volume} {414}},\ \bibinfo {pages} {509} (\bibinfo {year}
  {2001})}\BibitemShut {NoStop}%
\bibitem [{\citenamefont {Teichmann}\ \emph {et~al.}(2016)\citenamefont
  {Teichmann}, \citenamefont {Silva}, \citenamefont {Cousin}, \citenamefont
  {Hemmer},\ and\ \citenamefont {Biegert}}]{teichmann2016}%
  \BibitemOpen
  \bibfield  {author} {\bibinfo {author} {\bibfnamefont {S.~M.}\ \bibnamefont
  {Teichmann}}, \bibinfo {author} {\bibfnamefont {F.}~\bibnamefont {Silva}},
  \bibinfo {author} {\bibfnamefont {S.~L.}\ \bibnamefont {Cousin}}, \bibinfo
  {author} {\bibfnamefont {M.}~\bibnamefont {Hemmer}},\ and\ \bibinfo {author}
  {\bibfnamefont {J.}~\bibnamefont {Biegert}},\ }\bibfield  {title} {\bibinfo
  {title} {0.5-kev soft x-ray attosecond continua},\ }\href
  {https://doi.org/10.1038/ncomms11493} {\bibfield  {journal} {\bibinfo
  {journal} {Nat. Commun.}\ }\textbf {\bibinfo {volume} {7}},\ \bibinfo {pages}
  {11493} (\bibinfo {year} {2016})}\BibitemShut {NoStop}%
\bibitem [{\citenamefont {Gaumnitz}\ \emph {et~al.}(2017)\citenamefont
  {Gaumnitz}, \citenamefont {Jain}, \citenamefont {Pertot}, \citenamefont
  {Huppert}, \citenamefont {Jordan}, \citenamefont {Ardana-Lamas},\ and\
  \citenamefont {W\"{o}rner}}]{gaumnitz2017}%
  \BibitemOpen
  \bibfield  {author} {\bibinfo {author} {\bibfnamefont {T.}~\bibnamefont
  {Gaumnitz}}, \bibinfo {author} {\bibfnamefont {A.}~\bibnamefont {Jain}},
  \bibinfo {author} {\bibfnamefont {Y.}~\bibnamefont {Pertot}}, \bibinfo
  {author} {\bibfnamefont {M.}~\bibnamefont {Huppert}}, \bibinfo {author}
  {\bibfnamefont {I.}~\bibnamefont {Jordan}}, \bibinfo {author} {\bibfnamefont
  {F.}~\bibnamefont {Ardana-Lamas}},\ and\ \bibinfo {author} {\bibfnamefont
  {H.~J.}\ \bibnamefont {W\"{o}rner}},\ }\bibfield  {title} {\bibinfo {title}
  {Streaking of 43-attosecond soft-x-ray pulses generated by a passively
  cep-stable mid-infrared driver},\ }\href
  {https://doi.org/10.1364/OE.25.027506} {\bibfield  {journal} {\bibinfo
  {journal} {Opt. Express}\ }\textbf {\bibinfo {volume} {25}},\ \bibinfo
  {pages} {27506} (\bibinfo {year} {2017})}\BibitemShut {NoStop}%
\bibitem [{\citenamefont {Li}\ \emph {et~al.}(2017)\citenamefont {Li},
  \citenamefont {Ren}, \citenamefont {Yin}, \citenamefont {Zhao}, \citenamefont
  {Chew}, \citenamefont {Cheng}, \citenamefont {Cunningham}, \citenamefont
  {Wang}, \citenamefont {Hu}, \citenamefont {Wu}, \citenamefont {Chini},\ and\
  \citenamefont {Chang}}]{li2017}%
  \BibitemOpen
  \bibfield  {author} {\bibinfo {author} {\bibfnamefont {J.}~\bibnamefont
  {Li}}, \bibinfo {author} {\bibfnamefont {X.}~\bibnamefont {Ren}}, \bibinfo
  {author} {\bibfnamefont {Y.}~\bibnamefont {Yin}}, \bibinfo {author}
  {\bibfnamefont {K.}~\bibnamefont {Zhao}}, \bibinfo {author} {\bibfnamefont
  {A.}~\bibnamefont {Chew}}, \bibinfo {author} {\bibfnamefont {Y.}~\bibnamefont
  {Cheng}}, \bibinfo {author} {\bibfnamefont {E.}~\bibnamefont {Cunningham}},
  \bibinfo {author} {\bibfnamefont {Y.}~\bibnamefont {Wang}}, \bibinfo {author}
  {\bibfnamefont {S.}~\bibnamefont {Hu}}, \bibinfo {author} {\bibfnamefont
  {Y.}~\bibnamefont {Wu}}, \bibinfo {author} {\bibfnamefont {M.}~\bibnamefont
  {Chini}},\ and\ \bibinfo {author} {\bibfnamefont {Z.}~\bibnamefont {Chang}},\
  }\bibfield  {title} {\bibinfo {title} {53-attosecond x-ray pulses reach the
  carbon k-edge},\ }\href {https://doi.org/10.1038/s41467-017-00321-0}
  {\bibfield  {journal} {\bibinfo  {journal} {Nat. Commun.}\ }\textbf {\bibinfo
  {volume} {8}},\ \bibinfo {pages} {186} (\bibinfo {year} {2017})}\BibitemShut
  {NoStop}%
\bibitem [{\citenamefont {Duris}\ \emph {et~al.}(2020)\citenamefont {Duris},
  \citenamefont {Li}, \citenamefont {Driver}, \citenamefont {Champenois},
  \citenamefont {MacArthur}, \citenamefont {Lutman}, \citenamefont {Zhang},
  \citenamefont {Rosenberger}, \citenamefont {Aldrich}, \citenamefont {Coffee},
  \citenamefont {Coslovich}, \citenamefont {Decker}, \citenamefont {Glownia},
  \citenamefont {Hartmann}, \citenamefont {Helml}, \citenamefont {Kamalov},
  \citenamefont {Knurr}, \citenamefont {Krzywinski}, \citenamefont {Lin},
  \citenamefont {Marangos}, \citenamefont {Nantel}, \citenamefont {Natan},
  \citenamefont {O'Neal}, \citenamefont {Shivaram}, \citenamefont {Walter},
  \citenamefont {Wang}, \citenamefont {Welch}, \citenamefont {Wolf},
  \citenamefont {Xu}, \citenamefont {Kling}, \citenamefont {Bucksbaum},
  \citenamefont {Zholents}, \citenamefont {Huang}, \citenamefont {Cryan},\ and\
  \citenamefont {Marinelli}}]{duris2020}%
  \BibitemOpen
  \bibfield  {author} {\bibinfo {author} {\bibfnamefont {J.}~\bibnamefont
  {Duris}}, \bibinfo {author} {\bibfnamefont {S.}~\bibnamefont {Li}}, \bibinfo
  {author} {\bibfnamefont {T.}~\bibnamefont {Driver}}, \bibinfo {author}
  {\bibfnamefont {E.~G.}\ \bibnamefont {Champenois}}, \bibinfo {author}
  {\bibfnamefont {J.~P.}\ \bibnamefont {MacArthur}}, \bibinfo {author}
  {\bibfnamefont {A.~A.}\ \bibnamefont {Lutman}}, \bibinfo {author}
  {\bibfnamefont {Z.}~\bibnamefont {Zhang}}, \bibinfo {author} {\bibfnamefont
  {P.}~\bibnamefont {Rosenberger}}, \bibinfo {author} {\bibfnamefont {J.~W.}\
  \bibnamefont {Aldrich}}, \bibinfo {author} {\bibfnamefont {R.}~\bibnamefont
  {Coffee}}, \bibinfo {author} {\bibfnamefont {G.}~\bibnamefont {Coslovich}},
  \bibinfo {author} {\bibfnamefont {F.-J.}\ \bibnamefont {Decker}}, \bibinfo
  {author} {\bibfnamefont {J.~M.}\ \bibnamefont {Glownia}}, \bibinfo {author}
  {\bibfnamefont {G.}~\bibnamefont {Hartmann}}, \bibinfo {author}
  {\bibfnamefont {W.}~\bibnamefont {Helml}}, \bibinfo {author} {\bibfnamefont
  {A.}~\bibnamefont {Kamalov}}, \bibinfo {author} {\bibfnamefont
  {J.}~\bibnamefont {Knurr}}, \bibinfo {author} {\bibfnamefont
  {J.}~\bibnamefont {Krzywinski}}, \bibinfo {author} {\bibfnamefont {M.-F.}\
  \bibnamefont {Lin}}, \bibinfo {author} {\bibfnamefont {J.~P.}\ \bibnamefont
  {Marangos}}, \bibinfo {author} {\bibfnamefont {M.}~\bibnamefont {Nantel}},
  \bibinfo {author} {\bibfnamefont {A.}~\bibnamefont {Natan}}, \bibinfo
  {author} {\bibfnamefont {J.~T.}\ \bibnamefont {O'Neal}}, \bibinfo {author}
  {\bibfnamefont {N.}~\bibnamefont {Shivaram}}, \bibinfo {author}
  {\bibfnamefont {P.}~\bibnamefont {Walter}}, \bibinfo {author} {\bibfnamefont
  {A.~L.}\ \bibnamefont {Wang}}, \bibinfo {author} {\bibfnamefont {J.~J.}\
  \bibnamefont {Welch}}, \bibinfo {author} {\bibfnamefont {T.~J.~A.}\
  \bibnamefont {Wolf}}, \bibinfo {author} {\bibfnamefont {J.~Z.}\ \bibnamefont
  {Xu}}, \bibinfo {author} {\bibfnamefont {M.~F.}\ \bibnamefont {Kling}},
  \bibinfo {author} {\bibfnamefont {P.~H.}\ \bibnamefont {Bucksbaum}}, \bibinfo
  {author} {\bibfnamefont {A.}~\bibnamefont {Zholents}}, \bibinfo {author}
  {\bibfnamefont {Z.}~\bibnamefont {Huang}}, \bibinfo {author} {\bibfnamefont
  {J.~P.}\ \bibnamefont {Cryan}},\ and\ \bibinfo {author} {\bibfnamefont
  {A.}~\bibnamefont {Marinelli}},\ }\bibfield  {title} {\bibinfo {title}
  {Tunable isolated attosecond x-ray pulses with gigawatt peak power from a
  free-electron laser},\ }\href {https://doi.org/10.1038/s41566-019-0549-5}
  {\bibfield  {journal} {\bibinfo  {journal} {Nat. Photonics}\ }\textbf
  {\bibinfo {volume} {14}},\ \bibinfo {pages} {30} (\bibinfo {year}
  {2020})}\BibitemShut {NoStop}%
\bibitem [{\citenamefont {Maroju}\ \emph {et~al.}(2020)\citenamefont {Maroju},
  \citenamefont {Grazioli}, \citenamefont {Di~Fraia}, \citenamefont {Moioli},
  \citenamefont {Ertel}, \citenamefont {Ahmadi}, \citenamefont {Plekan},
  \citenamefont {Finetti}, \citenamefont {Allaria}, \citenamefont {Giannessi},
  \citenamefont {De~Ninno}, \citenamefont {Spezzani}, \citenamefont {Penco},
  \citenamefont {Spampinati}, \citenamefont {Demidovich}, \citenamefont
  {Danailov}, \citenamefont {Borghes}, \citenamefont {Kourousias},
  \citenamefont {Sanches Dos~Reis}, \citenamefont {Bill{\'e}}, \citenamefont
  {Lutman}, \citenamefont {Squibb}, \citenamefont {Feifel}, \citenamefont
  {Carpeggiani}, \citenamefont {Reduzzi}, \citenamefont {Mazza}, \citenamefont
  {Meyer}, \citenamefont {Bengtsson}, \citenamefont {Ibrakovic}, \citenamefont
  {Simpson}, \citenamefont {Mauritsson}, \citenamefont {Csizmadia},
  \citenamefont {Dumergue}, \citenamefont {K{\"u}hn}, \citenamefont
  {Nandiga~Gopalakrishna}, \citenamefont {You}, \citenamefont {Ueda},
  \citenamefont {Labeye}, \citenamefont {B{\ae}kh{\o}j}, \citenamefont
  {Schafer}, \citenamefont {Gryzlova}, \citenamefont {Grum-Grzhimailo},
  \citenamefont {Prince}, \citenamefont {Callegari},\ and\ \citenamefont
  {Sansone}}]{maroju2020}%
  \BibitemOpen
  \bibfield  {author} {\bibinfo {author} {\bibfnamefont {P.~K.}\ \bibnamefont
  {Maroju}}, \bibinfo {author} {\bibfnamefont {C.}~\bibnamefont {Grazioli}},
  \bibinfo {author} {\bibfnamefont {M.}~\bibnamefont {Di~Fraia}}, \bibinfo
  {author} {\bibfnamefont {M.}~\bibnamefont {Moioli}}, \bibinfo {author}
  {\bibfnamefont {D.}~\bibnamefont {Ertel}}, \bibinfo {author} {\bibfnamefont
  {H.}~\bibnamefont {Ahmadi}}, \bibinfo {author} {\bibfnamefont
  {O.}~\bibnamefont {Plekan}}, \bibinfo {author} {\bibfnamefont
  {P.}~\bibnamefont {Finetti}}, \bibinfo {author} {\bibfnamefont
  {E.}~\bibnamefont {Allaria}}, \bibinfo {author} {\bibfnamefont
  {L.}~\bibnamefont {Giannessi}}, \bibinfo {author} {\bibfnamefont
  {G.}~\bibnamefont {De~Ninno}}, \bibinfo {author} {\bibfnamefont
  {C.}~\bibnamefont {Spezzani}}, \bibinfo {author} {\bibfnamefont
  {G.}~\bibnamefont {Penco}}, \bibinfo {author} {\bibfnamefont
  {S.}~\bibnamefont {Spampinati}}, \bibinfo {author} {\bibfnamefont
  {A.}~\bibnamefont {Demidovich}}, \bibinfo {author} {\bibfnamefont {M.~B.}\
  \bibnamefont {Danailov}}, \bibinfo {author} {\bibfnamefont {R.}~\bibnamefont
  {Borghes}}, \bibinfo {author} {\bibfnamefont {G.}~\bibnamefont {Kourousias}},
  \bibinfo {author} {\bibfnamefont {C.~E.}\ \bibnamefont {Sanches Dos~Reis}},
  \bibinfo {author} {\bibfnamefont {F.}~\bibnamefont {Bill{\'e}}}, \bibinfo
  {author} {\bibfnamefont {A.~A.}\ \bibnamefont {Lutman}}, \bibinfo {author}
  {\bibfnamefont {R.~J.}\ \bibnamefont {Squibb}}, \bibinfo {author}
  {\bibfnamefont {R.}~\bibnamefont {Feifel}}, \bibinfo {author} {\bibfnamefont
  {P.}~\bibnamefont {Carpeggiani}}, \bibinfo {author} {\bibfnamefont
  {M.}~\bibnamefont {Reduzzi}}, \bibinfo {author} {\bibfnamefont
  {T.}~\bibnamefont {Mazza}}, \bibinfo {author} {\bibfnamefont
  {M.}~\bibnamefont {Meyer}}, \bibinfo {author} {\bibfnamefont
  {S.}~\bibnamefont {Bengtsson}}, \bibinfo {author} {\bibfnamefont
  {N.}~\bibnamefont {Ibrakovic}}, \bibinfo {author} {\bibfnamefont {E.~R.}\
  \bibnamefont {Simpson}}, \bibinfo {author} {\bibfnamefont {J.}~\bibnamefont
  {Mauritsson}}, \bibinfo {author} {\bibfnamefont {T.}~\bibnamefont
  {Csizmadia}}, \bibinfo {author} {\bibfnamefont {M.}~\bibnamefont {Dumergue}},
  \bibinfo {author} {\bibfnamefont {S.}~\bibnamefont {K{\"u}hn}}, \bibinfo
  {author} {\bibfnamefont {H.}~\bibnamefont {Nandiga~Gopalakrishna}}, \bibinfo
  {author} {\bibfnamefont {D.}~\bibnamefont {You}}, \bibinfo {author}
  {\bibfnamefont {K.}~\bibnamefont {Ueda}}, \bibinfo {author} {\bibfnamefont
  {M.}~\bibnamefont {Labeye}}, \bibinfo {author} {\bibfnamefont {J.~E.}\
  \bibnamefont {B{\ae}kh{\o}j}}, \bibinfo {author} {\bibfnamefont {K.~J.}\
  \bibnamefont {Schafer}}, \bibinfo {author} {\bibfnamefont {E.~V.}\
  \bibnamefont {Gryzlova}}, \bibinfo {author} {\bibfnamefont {A.~N.}\
  \bibnamefont {Grum-Grzhimailo}}, \bibinfo {author} {\bibfnamefont {K.~C.}\
  \bibnamefont {Prince}}, \bibinfo {author} {\bibfnamefont {C.}~\bibnamefont
  {Callegari}},\ and\ \bibinfo {author} {\bibfnamefont {G.}~\bibnamefont
  {Sansone}},\ }\bibfield  {title} {\bibinfo {title} {Attosecond pulse shaping
  using a seeded free-electron laser},\ }\href
  {https://doi.org/10.1038/s41586-020-2005-6} {\bibfield  {journal} {\bibinfo
  {journal} {Nature}\ }\textbf {\bibinfo {volume} {578}},\ \bibinfo {pages}
  {386} (\bibinfo {year} {2020})}\BibitemShut {NoStop}%
\bibitem [{\citenamefont {Franz}\ \emph {et~al.}(2024)\citenamefont {Franz},
  \citenamefont {Li}, \citenamefont {Driver}, \citenamefont {Robles},
  \citenamefont {Cesar}, \citenamefont {Isele}, \citenamefont {Guo},
  \citenamefont {Wang}, \citenamefont {Duris}, \citenamefont {Larsen},
  \citenamefont {Glownia}, \citenamefont {Cheng}, \citenamefont {Hoffmann},
  \citenamefont {Li}, \citenamefont {Lin}, \citenamefont {Kamalov},
  \citenamefont {Obaid}, \citenamefont {Summers}, \citenamefont {Sudar},
  \citenamefont {Thierstein}, \citenamefont {Zhang}, \citenamefont {Kling},
  \citenamefont {Huang}, \citenamefont {Cryan},\ and\ \citenamefont
  {Marinelli}}]{franz2024}%
  \BibitemOpen
  \bibfield  {author} {\bibinfo {author} {\bibfnamefont {P.}~\bibnamefont
  {Franz}}, \bibinfo {author} {\bibfnamefont {S.}~\bibnamefont {Li}}, \bibinfo
  {author} {\bibfnamefont {T.}~\bibnamefont {Driver}}, \bibinfo {author}
  {\bibfnamefont {R.~R.}\ \bibnamefont {Robles}}, \bibinfo {author}
  {\bibfnamefont {D.}~\bibnamefont {Cesar}}, \bibinfo {author} {\bibfnamefont
  {E.}~\bibnamefont {Isele}}, \bibinfo {author} {\bibfnamefont
  {Z.}~\bibnamefont {Guo}}, \bibinfo {author} {\bibfnamefont {J.}~\bibnamefont
  {Wang}}, \bibinfo {author} {\bibfnamefont {J.~P.}\ \bibnamefont {Duris}},
  \bibinfo {author} {\bibfnamefont {K.}~\bibnamefont {Larsen}}, \bibinfo
  {author} {\bibfnamefont {J.~M.}\ \bibnamefont {Glownia}}, \bibinfo {author}
  {\bibfnamefont {X.}~\bibnamefont {Cheng}}, \bibinfo {author} {\bibfnamefont
  {M.~C.}\ \bibnamefont {Hoffmann}}, \bibinfo {author} {\bibfnamefont
  {X.}~\bibnamefont {Li}}, \bibinfo {author} {\bibfnamefont {M.-F.}\
  \bibnamefont {Lin}}, \bibinfo {author} {\bibfnamefont {A.}~\bibnamefont
  {Kamalov}}, \bibinfo {author} {\bibfnamefont {R.}~\bibnamefont {Obaid}},
  \bibinfo {author} {\bibfnamefont {A.}~\bibnamefont {Summers}}, \bibinfo
  {author} {\bibfnamefont {N.}~\bibnamefont {Sudar}}, \bibinfo {author}
  {\bibfnamefont {E.}~\bibnamefont {Thierstein}}, \bibinfo {author}
  {\bibfnamefont {Z.}~\bibnamefont {Zhang}}, \bibinfo {author} {\bibfnamefont
  {M.~F.}\ \bibnamefont {Kling}}, \bibinfo {author} {\bibfnamefont
  {Z.}~\bibnamefont {Huang}}, \bibinfo {author} {\bibfnamefont {J.~P.}\
  \bibnamefont {Cryan}},\ and\ \bibinfo {author} {\bibfnamefont
  {A.}~\bibnamefont {Marinelli}},\ }\bibfield  {title} {\bibinfo {title}
  {Terawatt-scale attosecond x-ray pulses from a cascaded superradiant
  free-electron laser},\ }\href {https://doi.org/10.1038/s41566-024-01427-w}
  {\bibfield  {journal} {\bibinfo  {journal} {Nat. Photonics}\ }\textbf
  {\bibinfo {volume} {18}},\ \bibinfo {pages} {698} (\bibinfo {year}
  {2024})}\BibitemShut {NoStop}%
\bibitem [{\citenamefont {Huang}\ and\ \citenamefont {Kim}(2007)}]{huang2007}%
  \BibitemOpen
  \bibfield  {author} {\bibinfo {author} {\bibfnamefont {Z.}~\bibnamefont
  {Huang}}\ and\ \bibinfo {author} {\bibfnamefont {K.-J.}\ \bibnamefont
  {Kim}},\ }\bibfield  {title} {\bibinfo {title} {Review of x-ray free-electron
  laser theory},\ }\href {https://doi.org/10.1103/PhysRevSTAB.10.034801}
  {\bibfield  {journal} {\bibinfo  {journal} {Phys. Rev. ST Accel. Beams}\
  }\textbf {\bibinfo {volume} {10}},\ \bibinfo {pages} {034801} (\bibinfo
  {year} {2007})}\BibitemShut {NoStop}%
\bibitem [{\citenamefont {Schm{\"u}ser}\ \emph {et~al.}(2009)\citenamefont
  {Schm{\"u}ser}, \citenamefont {Dohlus}, \citenamefont {Rossbach},\ and\
  \citenamefont {Behrens}}]{schmueser2009}%
  \BibitemOpen
  \bibfield  {author} {\bibinfo {author} {\bibfnamefont {P.}~\bibnamefont
  {Schm{\"u}ser}}, \bibinfo {author} {\bibfnamefont {M.}~\bibnamefont
  {Dohlus}}, \bibinfo {author} {\bibfnamefont {J.}~\bibnamefont {Rossbach}},\
  and\ \bibinfo {author} {\bibfnamefont {C.}~\bibnamefont {Behrens}},\ }\href
  {https://doi.org/10.1007/978-3-540-79572-8} {\emph {\bibinfo {title}
  {Ultraviolet and Soft X-Ray Free-Electron Lasers}}},\ \bibinfo {edition}
  {2nd}\ ed.\ (\bibinfo  {publisher} {Springer Berlin Heidelberg},\ \bibinfo
  {year} {2009})\BibitemShut {NoStop}%
\bibitem [{\citenamefont {Emma}\ \emph {et~al.}(2010)\citenamefont {Emma},
  \citenamefont {Akre}, \citenamefont {Arthur}, \citenamefont {Bionta},
  \citenamefont {Bostedt}, \citenamefont {Bozek}, \citenamefont {Brachmann},
  \citenamefont {Bucksbaum}, \citenamefont {Coffee}, \citenamefont {Decker},
  \citenamefont {Ding}, \citenamefont {Dowell}, \citenamefont {Edstrom},
  \citenamefont {Fisher}, \citenamefont {Frisch}, \citenamefont {Gilevich},
  \citenamefont {Hastings}, \citenamefont {Hays}, \citenamefont {Hering},
  \citenamefont {Huang}, \citenamefont {Iverson}, \citenamefont {Loos},
  \citenamefont {Messerschmidt}, \citenamefont {Miahnahri}, \citenamefont
  {Moeller}, \citenamefont {Nuhn}, \citenamefont {Pile}, \citenamefont
  {Ratner}, \citenamefont {Rzepiela}, \citenamefont {Schultz}, \citenamefont
  {Smith}, \citenamefont {Stefan}, \citenamefont {Tompkins}, \citenamefont
  {Turner}, \citenamefont {Welch}, \citenamefont {White}, \citenamefont {Wu},
  \citenamefont {Yocky},\ and\ \citenamefont {Galayda}}]{emma2010}%
  \BibitemOpen
  \bibfield  {author} {\bibinfo {author} {\bibfnamefont {P.}~\bibnamefont
  {Emma}}, \bibinfo {author} {\bibfnamefont {R.}~\bibnamefont {Akre}}, \bibinfo
  {author} {\bibfnamefont {J.}~\bibnamefont {Arthur}}, \bibinfo {author}
  {\bibfnamefont {R.}~\bibnamefont {Bionta}}, \bibinfo {author} {\bibfnamefont
  {C.}~\bibnamefont {Bostedt}}, \bibinfo {author} {\bibfnamefont
  {J.}~\bibnamefont {Bozek}}, \bibinfo {author} {\bibfnamefont
  {A.}~\bibnamefont {Brachmann}}, \bibinfo {author} {\bibfnamefont
  {P.}~\bibnamefont {Bucksbaum}}, \bibinfo {author} {\bibfnamefont
  {R.}~\bibnamefont {Coffee}}, \bibinfo {author} {\bibfnamefont {F.-J.}\
  \bibnamefont {Decker}}, \bibinfo {author} {\bibfnamefont {Y.}~\bibnamefont
  {Ding}}, \bibinfo {author} {\bibfnamefont {D.}~\bibnamefont {Dowell}},
  \bibinfo {author} {\bibfnamefont {S.}~\bibnamefont {Edstrom}}, \bibinfo
  {author} {\bibfnamefont {A.}~\bibnamefont {Fisher}}, \bibinfo {author}
  {\bibfnamefont {J.}~\bibnamefont {Frisch}}, \bibinfo {author} {\bibfnamefont
  {S.}~\bibnamefont {Gilevich}}, \bibinfo {author} {\bibfnamefont
  {J.}~\bibnamefont {Hastings}}, \bibinfo {author} {\bibfnamefont
  {G.}~\bibnamefont {Hays}}, \bibinfo {author} {\bibfnamefont {P.}~\bibnamefont
  {Hering}}, \bibinfo {author} {\bibfnamefont {Z.}~\bibnamefont {Huang}},
  \bibinfo {author} {\bibfnamefont {R.}~\bibnamefont {Iverson}}, \bibinfo
  {author} {\bibfnamefont {H.}~\bibnamefont {Loos}}, \bibinfo {author}
  {\bibfnamefont {M.}~\bibnamefont {Messerschmidt}}, \bibinfo {author}
  {\bibfnamefont {A.}~\bibnamefont {Miahnahri}}, \bibinfo {author}
  {\bibfnamefont {S.}~\bibnamefont {Moeller}}, \bibinfo {author} {\bibfnamefont
  {H.-D.}\ \bibnamefont {Nuhn}}, \bibinfo {author} {\bibfnamefont
  {G.}~\bibnamefont {Pile}}, \bibinfo {author} {\bibfnamefont {D.}~\bibnamefont
  {Ratner}}, \bibinfo {author} {\bibfnamefont {J.}~\bibnamefont {Rzepiela}},
  \bibinfo {author} {\bibfnamefont {D.}~\bibnamefont {Schultz}}, \bibinfo
  {author} {\bibfnamefont {T.}~\bibnamefont {Smith}}, \bibinfo {author}
  {\bibfnamefont {P.}~\bibnamefont {Stefan}}, \bibinfo {author} {\bibfnamefont
  {H.}~\bibnamefont {Tompkins}}, \bibinfo {author} {\bibfnamefont
  {J.}~\bibnamefont {Turner}}, \bibinfo {author} {\bibfnamefont
  {J.}~\bibnamefont {Welch}}, \bibinfo {author} {\bibfnamefont
  {W.}~\bibnamefont {White}}, \bibinfo {author} {\bibfnamefont
  {J.}~\bibnamefont {Wu}}, \bibinfo {author} {\bibfnamefont {G.}~\bibnamefont
  {Yocky}},\ and\ \bibinfo {author} {\bibfnamefont {J.}~\bibnamefont
  {Galayda}},\ }\bibfield  {title} {\bibinfo {title} {First lasing and
  operation of an {\aa}ngstrom-wavelength free-electron laser},\ }\href
  {https://doi.org/10.1038/nphoton.2010.176} {\bibfield  {journal} {\bibinfo
  {journal} {Nat. Photonics}\ }\textbf {\bibinfo {volume} {4}},\ \bibinfo
  {pages} {641} (\bibinfo {year} {2010})}\BibitemShut {NoStop}%
\bibitem [{\citenamefont {Allaria}\ \emph {et~al.}(2012)\citenamefont
  {Allaria}, \citenamefont {Appio}, \citenamefont {Badano}, \citenamefont
  {Barletta}, \citenamefont {Bassanese}, \citenamefont {Biedron}, \citenamefont
  {Borga}, \citenamefont {Busetto}, \citenamefont {Castronovo}, \citenamefont
  {Cinquegrana}, \citenamefont {Cleva}, \citenamefont {Cocco}, \citenamefont
  {Cornacchia}, \citenamefont {Craievich}, \citenamefont {Cudin}, \citenamefont
  {D'Auria}, \citenamefont {Dal~Forno}, \citenamefont {Danailov}, \citenamefont
  {De~Monte}, \citenamefont {De~Ninno}, \citenamefont {Delgiusto},
  \citenamefont {Demidovich}, \citenamefont {Di~Mitri}, \citenamefont
  {Diviacco}, \citenamefont {Fabris}, \citenamefont {Fabris}, \citenamefont
  {Fawley}, \citenamefont {Ferianis}, \citenamefont {Ferrari}, \citenamefont
  {Ferry}, \citenamefont {Froehlich}, \citenamefont {Furlan}, \citenamefont
  {Gaio}, \citenamefont {Gelmetti}, \citenamefont {Giannessi}, \citenamefont
  {Giannini}, \citenamefont {Gobessi}, \citenamefont {Ivanov}, \citenamefont
  {Karantzoulis}, \citenamefont {Lonza}, \citenamefont {Lutman}, \citenamefont
  {Mahieu}, \citenamefont {Milloch}, \citenamefont {Milton}, \citenamefont
  {Musardo}, \citenamefont {Nikolov}, \citenamefont {Noe}, \citenamefont
  {Parmigiani}, \citenamefont {Penco}, \citenamefont {Petronio}, \citenamefont
  {Pivetta}, \citenamefont {Predonzani}, \citenamefont {Rossi}, \citenamefont
  {Rumiz}, \citenamefont {Salom}, \citenamefont {Scafuri}, \citenamefont
  {Serpico}, \citenamefont {Sigalotti}, \citenamefont {Spampinati},
  \citenamefont {Spezzani}, \citenamefont {Svandrlik}, \citenamefont {Svetina},
  \citenamefont {Tazzari}, \citenamefont {Trovo}, \citenamefont {Umer},
  \citenamefont {Vascotto}, \citenamefont {Veronese}, \citenamefont
  {Visintini}, \citenamefont {Zaccaria}, \citenamefont {Zangrando},\ and\
  \citenamefont {Zangrando}}]{allaria2012}%
  \BibitemOpen
  \bibfield  {author} {\bibinfo {author} {\bibfnamefont {E.}~\bibnamefont
  {Allaria}}, \bibinfo {author} {\bibfnamefont {R.}~\bibnamefont {Appio}},
  \bibinfo {author} {\bibfnamefont {L.}~\bibnamefont {Badano}}, \bibinfo
  {author} {\bibfnamefont {W.~A.}\ \bibnamefont {Barletta}}, \bibinfo {author}
  {\bibfnamefont {S.}~\bibnamefont {Bassanese}}, \bibinfo {author}
  {\bibfnamefont {S.~G.}\ \bibnamefont {Biedron}}, \bibinfo {author}
  {\bibfnamefont {A.}~\bibnamefont {Borga}}, \bibinfo {author} {\bibfnamefont
  {E.}~\bibnamefont {Busetto}}, \bibinfo {author} {\bibfnamefont
  {D.}~\bibnamefont {Castronovo}}, \bibinfo {author} {\bibfnamefont
  {P.}~\bibnamefont {Cinquegrana}}, \bibinfo {author} {\bibfnamefont
  {S.}~\bibnamefont {Cleva}}, \bibinfo {author} {\bibfnamefont
  {D.}~\bibnamefont {Cocco}}, \bibinfo {author} {\bibfnamefont
  {M.}~\bibnamefont {Cornacchia}}, \bibinfo {author} {\bibfnamefont
  {P.}~\bibnamefont {Craievich}}, \bibinfo {author} {\bibfnamefont
  {I.}~\bibnamefont {Cudin}}, \bibinfo {author} {\bibfnamefont
  {G.}~\bibnamefont {D'Auria}}, \bibinfo {author} {\bibfnamefont
  {M.}~\bibnamefont {Dal~Forno}}, \bibinfo {author} {\bibfnamefont {M.~B.}\
  \bibnamefont {Danailov}}, \bibinfo {author} {\bibfnamefont {R.}~\bibnamefont
  {De~Monte}}, \bibinfo {author} {\bibfnamefont {G.}~\bibnamefont {De~Ninno}},
  \bibinfo {author} {\bibfnamefont {P.}~\bibnamefont {Delgiusto}}, \bibinfo
  {author} {\bibfnamefont {A.}~\bibnamefont {Demidovich}}, \bibinfo {author}
  {\bibfnamefont {S.}~\bibnamefont {Di~Mitri}}, \bibinfo {author}
  {\bibfnamefont {B.}~\bibnamefont {Diviacco}}, \bibinfo {author}
  {\bibfnamefont {A.}~\bibnamefont {Fabris}}, \bibinfo {author} {\bibfnamefont
  {R.}~\bibnamefont {Fabris}}, \bibinfo {author} {\bibfnamefont
  {W.}~\bibnamefont {Fawley}}, \bibinfo {author} {\bibfnamefont
  {M.}~\bibnamefont {Ferianis}}, \bibinfo {author} {\bibfnamefont
  {E.}~\bibnamefont {Ferrari}}, \bibinfo {author} {\bibfnamefont
  {S.}~\bibnamefont {Ferry}}, \bibinfo {author} {\bibfnamefont
  {L.}~\bibnamefont {Froehlich}}, \bibinfo {author} {\bibfnamefont
  {P.}~\bibnamefont {Furlan}}, \bibinfo {author} {\bibfnamefont
  {G.}~\bibnamefont {Gaio}}, \bibinfo {author} {\bibfnamefont {F.}~\bibnamefont
  {Gelmetti}}, \bibinfo {author} {\bibfnamefont {L.}~\bibnamefont {Giannessi}},
  \bibinfo {author} {\bibfnamefont {M.}~\bibnamefont {Giannini}}, \bibinfo
  {author} {\bibfnamefont {R.}~\bibnamefont {Gobessi}}, \bibinfo {author}
  {\bibfnamefont {R.}~\bibnamefont {Ivanov}}, \bibinfo {author} {\bibfnamefont
  {E.}~\bibnamefont {Karantzoulis}}, \bibinfo {author} {\bibfnamefont
  {M.}~\bibnamefont {Lonza}}, \bibinfo {author} {\bibfnamefont
  {A.}~\bibnamefont {Lutman}}, \bibinfo {author} {\bibfnamefont
  {B.}~\bibnamefont {Mahieu}}, \bibinfo {author} {\bibfnamefont
  {M.}~\bibnamefont {Milloch}}, \bibinfo {author} {\bibfnamefont {S.~V.}\
  \bibnamefont {Milton}}, \bibinfo {author} {\bibfnamefont {M.}~\bibnamefont
  {Musardo}}, \bibinfo {author} {\bibfnamefont {I.}~\bibnamefont {Nikolov}},
  \bibinfo {author} {\bibfnamefont {S.}~\bibnamefont {Noe}}, \bibinfo {author}
  {\bibfnamefont {F.}~\bibnamefont {Parmigiani}}, \bibinfo {author}
  {\bibfnamefont {G.}~\bibnamefont {Penco}}, \bibinfo {author} {\bibfnamefont
  {M.}~\bibnamefont {Petronio}}, \bibinfo {author} {\bibfnamefont
  {L.}~\bibnamefont {Pivetta}}, \bibinfo {author} {\bibfnamefont
  {M.}~\bibnamefont {Predonzani}}, \bibinfo {author} {\bibfnamefont
  {F.}~\bibnamefont {Rossi}}, \bibinfo {author} {\bibfnamefont
  {L.}~\bibnamefont {Rumiz}}, \bibinfo {author} {\bibfnamefont
  {A.}~\bibnamefont {Salom}}, \bibinfo {author} {\bibfnamefont
  {C.}~\bibnamefont {Scafuri}}, \bibinfo {author} {\bibfnamefont
  {C.}~\bibnamefont {Serpico}}, \bibinfo {author} {\bibfnamefont
  {P.}~\bibnamefont {Sigalotti}}, \bibinfo {author} {\bibfnamefont
  {S.}~\bibnamefont {Spampinati}}, \bibinfo {author} {\bibfnamefont
  {C.}~\bibnamefont {Spezzani}}, \bibinfo {author} {\bibfnamefont
  {M.}~\bibnamefont {Svandrlik}}, \bibinfo {author} {\bibfnamefont
  {C.}~\bibnamefont {Svetina}}, \bibinfo {author} {\bibfnamefont
  {S.}~\bibnamefont {Tazzari}}, \bibinfo {author} {\bibfnamefont
  {M.}~\bibnamefont {Trovo}}, \bibinfo {author} {\bibfnamefont
  {R.}~\bibnamefont {Umer}}, \bibinfo {author} {\bibfnamefont {A.}~\bibnamefont
  {Vascotto}}, \bibinfo {author} {\bibfnamefont {M.}~\bibnamefont {Veronese}},
  \bibinfo {author} {\bibfnamefont {R.}~\bibnamefont {Visintini}}, \bibinfo
  {author} {\bibfnamefont {M.}~\bibnamefont {Zaccaria}}, \bibinfo {author}
  {\bibfnamefont {D.}~\bibnamefont {Zangrando}},\ and\ \bibinfo {author}
  {\bibfnamefont {M.}~\bibnamefont {Zangrando}},\ }\bibfield  {title} {\bibinfo
  {title} {Highly coherent and stable pulses from the fermi seeded
  free-electron laser in the extreme ultraviolet},\ }\href
  {https://doi.org/10.1038/nphoton.2012.233} {\bibfield  {journal} {\bibinfo
  {journal} {Nat. Photonics}\ }\textbf {\bibinfo {volume} {6}},\ \bibinfo
  {pages} {699} (\bibinfo {year} {2012})}\BibitemShut {NoStop}%
\bibitem [{\citenamefont {Prat}\ \emph {et~al.}(2020)\citenamefont {Prat},
  \citenamefont {Dijkstal}, \citenamefont {Ferrari},\ and\ \citenamefont
  {Reiche}}]{prat2020}%
  \BibitemOpen
  \bibfield  {author} {\bibinfo {author} {\bibfnamefont {E.}~\bibnamefont
  {Prat}}, \bibinfo {author} {\bibfnamefont {P.}~\bibnamefont {Dijkstal}},
  \bibinfo {author} {\bibfnamefont {E.}~\bibnamefont {Ferrari}},\ and\ \bibinfo
  {author} {\bibfnamefont {S.}~\bibnamefont {Reiche}},\ }\bibfield  {title}
  {\bibinfo {title} {Demonstration of large bandwidth hard x-ray free-electron
  laser pulses at swissfel},\ }\href
  {https://doi.org/10.1103/PhysRevLett.124.074801} {\bibfield  {journal}
  {\bibinfo  {journal} {Phys. Rev. Lett.}\ }\textbf {\bibinfo {volume} {124}},\
  \bibinfo {pages} {074801} (\bibinfo {year} {2020})}\BibitemShut {NoStop}%
\bibitem [{\citenamefont {Pompili}\ \emph {et~al.}(2022)\citenamefont
  {Pompili}, \citenamefont {Alesini}, \citenamefont {Anania}, \citenamefont
  {Arjmand}, \citenamefont {Behtouei}, \citenamefont {Bellaveglia},
  \citenamefont {Biagioni}, \citenamefont {Buonomo}, \citenamefont {Cardelli},
  \citenamefont {Carpanese}, \citenamefont {Chiadroni}, \citenamefont
  {Cianchi}, \citenamefont {Costa}, \citenamefont {Del~Dotto}, \citenamefont
  {Del~Giorno}, \citenamefont {Dipace}, \citenamefont {Doria}, \citenamefont
  {Filippi}, \citenamefont {Galletti}, \citenamefont {Giannessi}, \citenamefont
  {Giribono}, \citenamefont {Iovine}, \citenamefont {Lollo}, \citenamefont
  {Mostacci}, \citenamefont {Nguyen}, \citenamefont {Opromolla}, \citenamefont
  {Di~Palma}, \citenamefont {Pellegrino}, \citenamefont {Petralia},
  \citenamefont {Petrillo}, \citenamefont {Piersanti}, \citenamefont
  {Di~Pirro}, \citenamefont {Romeo}, \citenamefont {Rossi}, \citenamefont
  {Scifo}, \citenamefont {Selce}, \citenamefont {Shpakov}, \citenamefont
  {Stella}, \citenamefont {Vaccarezza}, \citenamefont {Villa}, \citenamefont
  {Zigler},\ and\ \citenamefont {Ferrario}}]{pompili2022}%
  \BibitemOpen
  \bibfield  {author} {\bibinfo {author} {\bibfnamefont {R.}~\bibnamefont
  {Pompili}}, \bibinfo {author} {\bibfnamefont {D.}~\bibnamefont {Alesini}},
  \bibinfo {author} {\bibfnamefont {M.~P.}\ \bibnamefont {Anania}}, \bibinfo
  {author} {\bibfnamefont {S.}~\bibnamefont {Arjmand}}, \bibinfo {author}
  {\bibfnamefont {M.}~\bibnamefont {Behtouei}}, \bibinfo {author}
  {\bibfnamefont {M.}~\bibnamefont {Bellaveglia}}, \bibinfo {author}
  {\bibfnamefont {A.}~\bibnamefont {Biagioni}}, \bibinfo {author}
  {\bibfnamefont {B.}~\bibnamefont {Buonomo}}, \bibinfo {author} {\bibfnamefont
  {F.}~\bibnamefont {Cardelli}}, \bibinfo {author} {\bibfnamefont
  {M.}~\bibnamefont {Carpanese}}, \bibinfo {author} {\bibfnamefont
  {E.}~\bibnamefont {Chiadroni}}, \bibinfo {author} {\bibfnamefont
  {A.}~\bibnamefont {Cianchi}}, \bibinfo {author} {\bibfnamefont
  {G.}~\bibnamefont {Costa}}, \bibinfo {author} {\bibfnamefont
  {A.}~\bibnamefont {Del~Dotto}}, \bibinfo {author} {\bibfnamefont
  {M.}~\bibnamefont {Del~Giorno}}, \bibinfo {author} {\bibfnamefont
  {F.}~\bibnamefont {Dipace}}, \bibinfo {author} {\bibfnamefont
  {A.}~\bibnamefont {Doria}}, \bibinfo {author} {\bibfnamefont
  {F.}~\bibnamefont {Filippi}}, \bibinfo {author} {\bibfnamefont
  {M.}~\bibnamefont {Galletti}}, \bibinfo {author} {\bibfnamefont
  {L.}~\bibnamefont {Giannessi}}, \bibinfo {author} {\bibfnamefont
  {A.}~\bibnamefont {Giribono}}, \bibinfo {author} {\bibfnamefont
  {P.}~\bibnamefont {Iovine}}, \bibinfo {author} {\bibfnamefont
  {V.}~\bibnamefont {Lollo}}, \bibinfo {author} {\bibfnamefont
  {A.}~\bibnamefont {Mostacci}}, \bibinfo {author} {\bibfnamefont
  {F.}~\bibnamefont {Nguyen}}, \bibinfo {author} {\bibfnamefont
  {M.}~\bibnamefont {Opromolla}}, \bibinfo {author} {\bibfnamefont
  {E.}~\bibnamefont {Di~Palma}}, \bibinfo {author} {\bibfnamefont
  {L.}~\bibnamefont {Pellegrino}}, \bibinfo {author} {\bibfnamefont
  {A.}~\bibnamefont {Petralia}}, \bibinfo {author} {\bibfnamefont
  {V.}~\bibnamefont {Petrillo}}, \bibinfo {author} {\bibfnamefont
  {L.}~\bibnamefont {Piersanti}}, \bibinfo {author} {\bibfnamefont
  {G.}~\bibnamefont {Di~Pirro}}, \bibinfo {author} {\bibfnamefont
  {S.}~\bibnamefont {Romeo}}, \bibinfo {author} {\bibfnamefont {A.~R.}\
  \bibnamefont {Rossi}}, \bibinfo {author} {\bibfnamefont {J.}~\bibnamefont
  {Scifo}}, \bibinfo {author} {\bibfnamefont {A.}~\bibnamefont {Selce}},
  \bibinfo {author} {\bibfnamefont {V.}~\bibnamefont {Shpakov}}, \bibinfo
  {author} {\bibfnamefont {A.}~\bibnamefont {Stella}}, \bibinfo {author}
  {\bibfnamefont {C.}~\bibnamefont {Vaccarezza}}, \bibinfo {author}
  {\bibfnamefont {F.}~\bibnamefont {Villa}}, \bibinfo {author} {\bibfnamefont
  {A.}~\bibnamefont {Zigler}},\ and\ \bibinfo {author} {\bibfnamefont
  {M.}~\bibnamefont {Ferrario}},\ }\bibfield  {title} {\bibinfo {title}
  {Free-electron lasing with compact beam-driven plasma wakefield
  accelerator},\ }\href {https://doi.org/10.1038/s41586-022-04589-1} {\bibfield
   {journal} {\bibinfo  {journal} {Nature}\ }\textbf {\bibinfo {volume}
  {605}},\ \bibinfo {pages} {659} (\bibinfo {year} {2022})}\BibitemShut
  {NoStop}%
\bibitem [{\citenamefont {Labat}\ \emph {et~al.}(2023)\citenamefont {Labat},
  \citenamefont {Cabada{\u{g}}}, \citenamefont {Ghaith}, \citenamefont {Irman},
  \citenamefont {Berlioux}, \citenamefont {Berteaud}, \citenamefont {Blache},
  \citenamefont {Bock}, \citenamefont {Bouvet}, \citenamefont {Briquez},
  \citenamefont {Chang}, \citenamefont {Corde}, \citenamefont {Debus},
  \citenamefont {De~Oliveira}, \citenamefont {Duval}, \citenamefont {Dietrich},
  \citenamefont {El~Ajjouri}, \citenamefont {Eisenmann}, \citenamefont
  {Gautier}, \citenamefont {Gebhardt}, \citenamefont {Grams}, \citenamefont
  {Helbig}, \citenamefont {Herbeaux}, \citenamefont {Hubert}, \citenamefont
  {Kitegi}, \citenamefont {Kononenko}, \citenamefont {Kuntzsch}, \citenamefont
  {LaBerge}, \citenamefont {L{\^e}}, \citenamefont {Leluan}, \citenamefont
  {Loulergue}, \citenamefont {Malka}, \citenamefont {Marteau}, \citenamefont
  {Guyen}, \citenamefont {Oumbarek-Espinos}, \citenamefont {Pausch},
  \citenamefont {Pereira}, \citenamefont {P{\"u}schel}, \citenamefont {Ricaud},
  \citenamefont {Rommeluere}, \citenamefont {Roussel}, \citenamefont
  {Rousseau}, \citenamefont {Sch{\"o}bel}, \citenamefont {Sebdaoui},
  \citenamefont {Steiniger}, \citenamefont {Tavakoli}, \citenamefont {Thaury},
  \citenamefont {Ufer}, \citenamefont {Vall{\'e}au}, \citenamefont
  {Vandenberghe}, \citenamefont {V{\'e}t{\'e}ran}, \citenamefont {Schramm},\
  and\ \citenamefont {Couprie}}]{labat2023}%
  \BibitemOpen
  \bibfield  {author} {\bibinfo {author} {\bibfnamefont {M.}~\bibnamefont
  {Labat}}, \bibinfo {author} {\bibfnamefont {J.~C.}\ \bibnamefont
  {Cabada{\u{g}}}}, \bibinfo {author} {\bibfnamefont {A.}~\bibnamefont
  {Ghaith}}, \bibinfo {author} {\bibfnamefont {A.}~\bibnamefont {Irman}},
  \bibinfo {author} {\bibfnamefont {A.}~\bibnamefont {Berlioux}}, \bibinfo
  {author} {\bibfnamefont {P.}~\bibnamefont {Berteaud}}, \bibinfo {author}
  {\bibfnamefont {F.}~\bibnamefont {Blache}}, \bibinfo {author} {\bibfnamefont
  {S.}~\bibnamefont {Bock}}, \bibinfo {author} {\bibfnamefont {F.}~\bibnamefont
  {Bouvet}}, \bibinfo {author} {\bibfnamefont {F.}~\bibnamefont {Briquez}},
  \bibinfo {author} {\bibfnamefont {Y.-Y.}\ \bibnamefont {Chang}}, \bibinfo
  {author} {\bibfnamefont {S.}~\bibnamefont {Corde}}, \bibinfo {author}
  {\bibfnamefont {A.}~\bibnamefont {Debus}}, \bibinfo {author} {\bibfnamefont
  {C.}~\bibnamefont {De~Oliveira}}, \bibinfo {author} {\bibfnamefont {J.-P.}\
  \bibnamefont {Duval}}, \bibinfo {author} {\bibfnamefont {Y.}~\bibnamefont
  {Dietrich}}, \bibinfo {author} {\bibfnamefont {M.}~\bibnamefont
  {El~Ajjouri}}, \bibinfo {author} {\bibfnamefont {C.}~\bibnamefont
  {Eisenmann}}, \bibinfo {author} {\bibfnamefont {J.}~\bibnamefont {Gautier}},
  \bibinfo {author} {\bibfnamefont {R.}~\bibnamefont {Gebhardt}}, \bibinfo
  {author} {\bibfnamefont {S.}~\bibnamefont {Grams}}, \bibinfo {author}
  {\bibfnamefont {U.}~\bibnamefont {Helbig}}, \bibinfo {author} {\bibfnamefont
  {C.}~\bibnamefont {Herbeaux}}, \bibinfo {author} {\bibfnamefont
  {N.}~\bibnamefont {Hubert}}, \bibinfo {author} {\bibfnamefont
  {C.}~\bibnamefont {Kitegi}}, \bibinfo {author} {\bibfnamefont
  {O.}~\bibnamefont {Kononenko}}, \bibinfo {author} {\bibfnamefont
  {M.}~\bibnamefont {Kuntzsch}}, \bibinfo {author} {\bibfnamefont
  {M.}~\bibnamefont {LaBerge}}, \bibinfo {author} {\bibfnamefont
  {S.}~\bibnamefont {L{\^e}}}, \bibinfo {author} {\bibfnamefont
  {B.}~\bibnamefont {Leluan}}, \bibinfo {author} {\bibfnamefont
  {A.}~\bibnamefont {Loulergue}}, \bibinfo {author} {\bibfnamefont
  {V.}~\bibnamefont {Malka}}, \bibinfo {author} {\bibfnamefont
  {F.}~\bibnamefont {Marteau}}, \bibinfo {author} {\bibfnamefont {M.~H.~N.}\
  \bibnamefont {Guyen}}, \bibinfo {author} {\bibfnamefont {D.}~\bibnamefont
  {Oumbarek-Espinos}}, \bibinfo {author} {\bibfnamefont {R.}~\bibnamefont
  {Pausch}}, \bibinfo {author} {\bibfnamefont {D.}~\bibnamefont {Pereira}},
  \bibinfo {author} {\bibfnamefont {T.}~\bibnamefont {P{\"u}schel}}, \bibinfo
  {author} {\bibfnamefont {J.-P.}\ \bibnamefont {Ricaud}}, \bibinfo {author}
  {\bibfnamefont {P.}~\bibnamefont {Rommeluere}}, \bibinfo {author}
  {\bibfnamefont {E.}~\bibnamefont {Roussel}}, \bibinfo {author} {\bibfnamefont
  {P.}~\bibnamefont {Rousseau}}, \bibinfo {author} {\bibfnamefont
  {S.}~\bibnamefont {Sch{\"o}bel}}, \bibinfo {author} {\bibfnamefont
  {M.}~\bibnamefont {Sebdaoui}}, \bibinfo {author} {\bibfnamefont
  {K.}~\bibnamefont {Steiniger}}, \bibinfo {author} {\bibfnamefont
  {K.}~\bibnamefont {Tavakoli}}, \bibinfo {author} {\bibfnamefont
  {C.}~\bibnamefont {Thaury}}, \bibinfo {author} {\bibfnamefont
  {P.}~\bibnamefont {Ufer}}, \bibinfo {author} {\bibfnamefont {M.}~\bibnamefont
  {Vall{\'e}au}}, \bibinfo {author} {\bibfnamefont {M.}~\bibnamefont
  {Vandenberghe}}, \bibinfo {author} {\bibfnamefont {J.}~\bibnamefont
  {V{\'e}t{\'e}ran}}, \bibinfo {author} {\bibfnamefont {U.}~\bibnamefont
  {Schramm}},\ and\ \bibinfo {author} {\bibfnamefont {M.-E.}\ \bibnamefont
  {Couprie}},\ }\bibfield  {title} {\bibinfo {title} {Seeded free-electron
  laser driven by a compact laser plasma accelerator},\ }\href
  {https://doi.org/10.1038/s41566-022-01104-w} {\bibfield  {journal} {\bibinfo
  {journal} {Nat. Photonics}\ }\textbf {\bibinfo {volume} {17}},\ \bibinfo
  {pages} {150} (\bibinfo {year} {2023})}\BibitemShut {NoStop}%
\bibitem [{\citenamefont {Habib}\ \emph {et~al.}(2023)\citenamefont {Habib},
  \citenamefont {Manahan}, \citenamefont {Scherkl}, \citenamefont {Heinemann},
  \citenamefont {Sutherland}, \citenamefont {Altuiri}, \citenamefont
  {Alotaibi}, \citenamefont {Litos}, \citenamefont {Cary}, \citenamefont
  {Raubenheimer}, \citenamefont {Hemsing}, \citenamefont {Hogan}, \citenamefont
  {Rosenzweig}, \citenamefont {Williams}, \citenamefont {McNeil},\ and\
  \citenamefont {Hidding}}]{habib2023}%
  \BibitemOpen
  \bibfield  {author} {\bibinfo {author} {\bibfnamefont {A.~F.}\ \bibnamefont
  {Habib}}, \bibinfo {author} {\bibfnamefont {G.~G.}\ \bibnamefont {Manahan}},
  \bibinfo {author} {\bibfnamefont {P.}~\bibnamefont {Scherkl}}, \bibinfo
  {author} {\bibfnamefont {T.}~\bibnamefont {Heinemann}}, \bibinfo {author}
  {\bibfnamefont {A.}~\bibnamefont {Sutherland}}, \bibinfo {author}
  {\bibfnamefont {R.}~\bibnamefont {Altuiri}}, \bibinfo {author} {\bibfnamefont
  {B.~M.}\ \bibnamefont {Alotaibi}}, \bibinfo {author} {\bibfnamefont
  {M.}~\bibnamefont {Litos}}, \bibinfo {author} {\bibfnamefont
  {J.}~\bibnamefont {Cary}}, \bibinfo {author} {\bibfnamefont {T.}~\bibnamefont
  {Raubenheimer}}, \bibinfo {author} {\bibfnamefont {E.}~\bibnamefont
  {Hemsing}}, \bibinfo {author} {\bibfnamefont {M.~J.}\ \bibnamefont {Hogan}},
  \bibinfo {author} {\bibfnamefont {J.~B.}\ \bibnamefont {Rosenzweig}},
  \bibinfo {author} {\bibfnamefont {P.~H.}\ \bibnamefont {Williams}}, \bibinfo
  {author} {\bibfnamefont {B.~W.~J.}\ \bibnamefont {McNeil}},\ and\ \bibinfo
  {author} {\bibfnamefont {B.}~\bibnamefont {Hidding}},\ }\bibfield  {title}
  {\bibinfo {title} {Attosecond-angstrom free-electron-laser towards the cold
  beam limit},\ }\href {https://doi.org/10.1038/s41467-023-36592-z} {\bibfield
  {journal} {\bibinfo  {journal} {Nat. Commun.}\ }\textbf {\bibinfo {volume}
  {14}},\ \bibinfo {pages} {1054} (\bibinfo {year} {2023})}\BibitemShut
  {NoStop}%
\bibitem [{\citenamefont {Gea-Banacloche}\ \emph {et~al.}(1987)\citenamefont
  {Gea-Banacloche}, \citenamefont {Moore}, \citenamefont {Schlicher},
  \citenamefont {Scully},\ and\ \citenamefont {Walther}}]{geabanacloche1987}%
  \BibitemOpen
  \bibfield  {author} {\bibinfo {author} {\bibfnamefont {J.}~\bibnamefont
  {Gea-Banacloche}}, \bibinfo {author} {\bibfnamefont {G.}~\bibnamefont
  {Moore}}, \bibinfo {author} {\bibfnamefont {R.}~\bibnamefont {Schlicher}},
  \bibinfo {author} {\bibfnamefont {M.}~\bibnamefont {Scully}},\ and\ \bibinfo
  {author} {\bibfnamefont {H.}~\bibnamefont {Walther}},\ }\bibfield  {title}
  {\bibinfo {title} {Soft x-ray free-electron laser with a laser undulator},\
  }\href {https://doi.org/10.1109/JQE.1987.1073559} {\bibfield  {journal}
  {\bibinfo  {journal} {IEEE J. Quantum Electron.}\ }\textbf {\bibinfo {volume}
  {23}},\ \bibinfo {pages} {1558} (\bibinfo {year} {1987})}\BibitemShut
  {NoStop}%
\bibitem [{\citenamefont {Gallardo}\ \emph {et~al.}(1988)\citenamefont
  {Gallardo}, \citenamefont {Fernow}, \citenamefont {Palmer},\ and\
  \citenamefont {Pellegrini}}]{gallardo1988}%
  \BibitemOpen
  \bibfield  {author} {\bibinfo {author} {\bibfnamefont {J.}~\bibnamefont
  {Gallardo}}, \bibinfo {author} {\bibfnamefont {R.}~\bibnamefont {Fernow}},
  \bibinfo {author} {\bibfnamefont {R.}~\bibnamefont {Palmer}},\ and\ \bibinfo
  {author} {\bibfnamefont {C.}~\bibnamefont {Pellegrini}},\ }\bibfield  {title}
  {\bibinfo {title} {Theory of a free-electron laser with a gaussian optical
  undulator},\ }\href {https://doi.org/10.1109/3.7085} {\bibfield  {journal}
  {\bibinfo  {journal} {IEEE J. Quantum Electron.}\ }\textbf {\bibinfo {volume}
  {24}},\ \bibinfo {pages} {1557} (\bibinfo {year} {1988})}\BibitemShut
  {NoStop}%
\bibitem [{\citenamefont {Steiniger}\ \emph {et~al.}(2014)\citenamefont
  {Steiniger}, \citenamefont {Bussmann}, \citenamefont {Pausch}, \citenamefont
  {Cowan}, \citenamefont {Irman}, \citenamefont {Jochmann}, \citenamefont
  {Sauerbrey}, \citenamefont {Schramm},\ and\ \citenamefont
  {Debus}}]{steiniger2014}%
  \BibitemOpen
  \bibfield  {author} {\bibinfo {author} {\bibfnamefont {K.}~\bibnamefont
  {Steiniger}}, \bibinfo {author} {\bibfnamefont {M.}~\bibnamefont {Bussmann}},
  \bibinfo {author} {\bibfnamefont {R.}~\bibnamefont {Pausch}}, \bibinfo
  {author} {\bibfnamefont {T.}~\bibnamefont {Cowan}}, \bibinfo {author}
  {\bibfnamefont {A.}~\bibnamefont {Irman}}, \bibinfo {author} {\bibfnamefont
  {A.}~\bibnamefont {Jochmann}}, \bibinfo {author} {\bibfnamefont
  {R.}~\bibnamefont {Sauerbrey}}, \bibinfo {author} {\bibfnamefont
  {U.}~\bibnamefont {Schramm}},\ and\ \bibinfo {author} {\bibfnamefont
  {A.}~\bibnamefont {Debus}},\ }\bibfield  {title} {\bibinfo {title} {Optical
  free-electron lasers with traveling-wave thomson-scattering},\ }\href
  {https://doi.org/10.1088/0953-4075/47/23/234011} {\bibfield  {journal}
  {\bibinfo  {journal} {J. Phys. B: At. Mol. Opt. Phys.}\ }\textbf {\bibinfo
  {volume} {47}},\ \bibinfo {pages} {234011} (\bibinfo {year}
  {2014})}\BibitemShut {NoStop}%
\bibitem [{\citenamefont {Steiniger}\ \emph {et~al.}(2019)\citenamefont
  {Steiniger}, \citenamefont {Albach}, \citenamefont {Bussmann}, \citenamefont
  {Loeser}, \citenamefont {Pausch}, \citenamefont {Röser}, \citenamefont
  {Schramm}, \citenamefont {Siebold},\ and\ \citenamefont
  {Debus}}]{steiniger2019}%
  \BibitemOpen
  \bibfield  {author} {\bibinfo {author} {\bibfnamefont {K.}~\bibnamefont
  {Steiniger}}, \bibinfo {author} {\bibfnamefont {D.}~\bibnamefont {Albach}},
  \bibinfo {author} {\bibfnamefont {M.}~\bibnamefont {Bussmann}}, \bibinfo
  {author} {\bibfnamefont {M.}~\bibnamefont {Loeser}}, \bibinfo {author}
  {\bibfnamefont {R.}~\bibnamefont {Pausch}}, \bibinfo {author} {\bibfnamefont
  {F.}~\bibnamefont {Röser}}, \bibinfo {author} {\bibfnamefont
  {U.}~\bibnamefont {Schramm}}, \bibinfo {author} {\bibfnamefont
  {M.}~\bibnamefont {Siebold}},\ and\ \bibinfo {author} {\bibfnamefont
  {A.}~\bibnamefont {Debus}},\ }\bibfield  {title} {\bibinfo {title} {Building
  an optical free-electron laser in the traveling-wave thomson-scattering
  geometry},\ }\href {https://doi.org/10.3389/fphy.2018.00155} {\bibfield
  {journal} {\bibinfo  {journal} {Front. Phys.}\ }\textbf {\bibinfo {volume}
  {6}} (\bibinfo {year} {2019})}\BibitemShut {NoStop}%
\bibitem [{\citenamefont {Sarachik}\ and\ \citenamefont
  {Schappert}(1970)}]{sarachik1970}%
  \BibitemOpen
  \bibfield  {author} {\bibinfo {author} {\bibfnamefont {E.~S.}\ \bibnamefont
  {Sarachik}}\ and\ \bibinfo {author} {\bibfnamefont {G.~T.}\ \bibnamefont
  {Schappert}},\ }\bibfield  {title} {\bibinfo {title} {Classical theory of the
  scattering of intense laser radiation by free electrons},\ }\href
  {https://doi.org/10.1103/PhysRevD.1.2738} {\bibfield  {journal} {\bibinfo
  {journal} {Phys. Rev. D}\ }\textbf {\bibinfo {volume} {1}},\ \bibinfo {pages}
  {2738} (\bibinfo {year} {1970})}\BibitemShut {NoStop}%
\bibitem [{\citenamefont {Salamin}\ and\ \citenamefont
  {Faisal}(1996)}]{salamin1996}%
  \BibitemOpen
  \bibfield  {author} {\bibinfo {author} {\bibfnamefont {Y.~I.}\ \bibnamefont
  {Salamin}}\ and\ \bibinfo {author} {\bibfnamefont {F.~H.~M.}\ \bibnamefont
  {Faisal}},\ }\bibfield  {title} {\bibinfo {title} {Harmonic generation by
  superintense light scattering from relativistic electrons},\ }\href
  {https://doi.org/10.1103/PhysRevA.54.4383} {\bibfield  {journal} {\bibinfo
  {journal} {Phys. Rev. A}\ }\textbf {\bibinfo {volume} {54}},\ \bibinfo
  {pages} {4383} (\bibinfo {year} {1996})}\BibitemShut {NoStop}%
\bibitem [{\citenamefont {Schreiber}\ and\ \citenamefont
  {Faatz}(2015)}]{schreiber2015}%
  \BibitemOpen
  \bibfield  {author} {\bibinfo {author} {\bibfnamefont {S.}~\bibnamefont
  {Schreiber}}\ and\ \bibinfo {author} {\bibfnamefont {B.}~\bibnamefont
  {Faatz}},\ }\bibfield  {title} {\bibinfo {title} {The free-electron laser
  flash},\ }\href {https://doi.org/10.1017/hpl.2015.16} {\bibfield  {journal}
  {\bibinfo  {journal} {High Power Laser Sci. Eng.}\ }\textbf {\bibinfo
  {volume} {3}},\ \bibinfo {pages} {e20} (\bibinfo {year} {2015})}\BibitemShut
  {NoStop}%
\bibitem [{\citenamefont {{Sarri}}\ \emph {et~al.}(2015)\citenamefont
  {{Sarri}}, \citenamefont {{Poder}}, \citenamefont {{Cole}}, \citenamefont
  {{Schumaker}}, \citenamefont {{di Piazza}}, \citenamefont {{Reville}},
  \citenamefont {{Dzelzainis}}, \citenamefont {{Doria}}, \citenamefont
  {{Gizzi}}, \citenamefont {{Grittani}}, \citenamefont {{Kar}}, \citenamefont
  {{Keitel}}, \citenamefont {{Krushelnick}}, \citenamefont {{Kuschel}},
  \citenamefont {{Mangles}}, \citenamefont {{Najmudin}}, \citenamefont
  {{Shukla}}, \citenamefont {{Silva}}, \citenamefont {{Symes}}, \citenamefont
  {{Thomas}}, \citenamefont {{Vargas}}, \citenamefont {{Vieira}},\ and\
  \citenamefont {{Zepf}}}]{sarri2015}%
  \BibitemOpen
  \bibfield  {author} {\bibinfo {author} {\bibfnamefont {G.}~\bibnamefont
  {{Sarri}}}, \bibinfo {author} {\bibfnamefont {K.}~\bibnamefont {{Poder}}},
  \bibinfo {author} {\bibfnamefont {J.~M.}\ \bibnamefont {{Cole}}}, \bibinfo
  {author} {\bibfnamefont {W.}~\bibnamefont {{Schumaker}}}, \bibinfo {author}
  {\bibfnamefont {A.}~\bibnamefont {{di Piazza}}}, \bibinfo {author}
  {\bibfnamefont {B.}~\bibnamefont {{Reville}}}, \bibinfo {author}
  {\bibfnamefont {T.}~\bibnamefont {{Dzelzainis}}}, \bibinfo {author}
  {\bibfnamefont {D.}~\bibnamefont {{Doria}}}, \bibinfo {author} {\bibfnamefont
  {L.~A.}\ \bibnamefont {{Gizzi}}}, \bibinfo {author} {\bibfnamefont
  {G.}~\bibnamefont {{Grittani}}}, \bibinfo {author} {\bibfnamefont
  {S.}~\bibnamefont {{Kar}}}, \bibinfo {author} {\bibfnamefont {C.~H.}\
  \bibnamefont {{Keitel}}}, \bibinfo {author} {\bibfnamefont {K.}~\bibnamefont
  {{Krushelnick}}}, \bibinfo {author} {\bibfnamefont {S.}~\bibnamefont
  {{Kuschel}}}, \bibinfo {author} {\bibfnamefont {S.~P.~D.}\ \bibnamefont
  {{Mangles}}}, \bibinfo {author} {\bibfnamefont {Z.}~\bibnamefont
  {{Najmudin}}}, \bibinfo {author} {\bibfnamefont {N.}~\bibnamefont
  {{Shukla}}}, \bibinfo {author} {\bibfnamefont {L.~O.}\ \bibnamefont
  {{Silva}}}, \bibinfo {author} {\bibfnamefont {D.}~\bibnamefont {{Symes}}},
  \bibinfo {author} {\bibfnamefont {A.~G.~R.}\ \bibnamefont {{Thomas}}},
  \bibinfo {author} {\bibfnamefont {M.}~\bibnamefont {{Vargas}}}, \bibinfo
  {author} {\bibfnamefont {J.}~\bibnamefont {{Vieira}}},\ and\ \bibinfo
  {author} {\bibfnamefont {M.}~\bibnamefont {{Zepf}}},\ }\bibfield  {title}
  {\bibinfo {title} {{Generation of neutral and high-density electron-positron
  pair plasmas in the laboratory}},\ }\href
  {https://doi.org/10.1038/ncomms7747} {\bibfield  {journal} {\bibinfo
  {journal} {Nat. Commun.}\ }\textbf {\bibinfo {volume} {6}},\ \bibinfo {eid}
  {6747} (\bibinfo {year} {2015})}\BibitemShut {NoStop}%
\bibitem [{\citenamefont {Chen}\ \emph {et~al.}(2015)\citenamefont {Chen},
  \citenamefont {Fiuza}, \citenamefont {Link}, \citenamefont {Hazi},
  \citenamefont {Hill}, \citenamefont {Hoarty}, \citenamefont {James},
  \citenamefont {Kerr}, \citenamefont {Meyerhofer}, \citenamefont {Myatt},
  \citenamefont {Park}, \citenamefont {Sentoku},\ and\ \citenamefont
  {Williams}}]{chen2015}%
  \BibitemOpen
  \bibfield  {author} {\bibinfo {author} {\bibfnamefont {H.}~\bibnamefont
  {Chen}}, \bibinfo {author} {\bibfnamefont {F.}~\bibnamefont {Fiuza}},
  \bibinfo {author} {\bibfnamefont {A.}~\bibnamefont {Link}}, \bibinfo {author}
  {\bibfnamefont {A.}~\bibnamefont {Hazi}}, \bibinfo {author} {\bibfnamefont
  {M.}~\bibnamefont {Hill}}, \bibinfo {author} {\bibfnamefont {D.}~\bibnamefont
  {Hoarty}}, \bibinfo {author} {\bibfnamefont {S.}~\bibnamefont {James}},
  \bibinfo {author} {\bibfnamefont {S.}~\bibnamefont {Kerr}}, \bibinfo {author}
  {\bibfnamefont {D.~D.}\ \bibnamefont {Meyerhofer}}, \bibinfo {author}
  {\bibfnamefont {J.}~\bibnamefont {Myatt}}, \bibinfo {author} {\bibfnamefont
  {J.}~\bibnamefont {Park}}, \bibinfo {author} {\bibfnamefont {Y.}~\bibnamefont
  {Sentoku}},\ and\ \bibinfo {author} {\bibfnamefont {G.~J.}\ \bibnamefont
  {Williams}},\ }\bibfield  {title} {\bibinfo {title} {Scaling the yield of
  laser-driven electron-positron jets to laboratory astrophysical
  applications},\ }\href {https://doi.org/10.1103/PhysRevLett.114.215001}
  {\bibfield  {journal} {\bibinfo  {journal} {Phys. Rev. Lett.}\ }\textbf
  {\bibinfo {volume} {114}},\ \bibinfo {pages} {215001} (\bibinfo {year}
  {2015})}\BibitemShut {NoStop}%
\bibitem [{\citenamefont {Liang}\ \emph {et~al.}(2015)\citenamefont {Liang},
  \citenamefont {Clarke}, \citenamefont {Henderson}, \citenamefont {Fu},
  \citenamefont {Lo}, \citenamefont {Taylor}, \citenamefont {Chaguine},
  \citenamefont {Zhou}, \citenamefont {Hua}, \citenamefont {Cen}, \citenamefont
  {Wang}, \citenamefont {Kao}, \citenamefont {Hasson}, \citenamefont {Dyer},
  \citenamefont {Serratto}, \citenamefont {Riley}, \citenamefont {Donovan},\
  and\ \citenamefont {Ditmire}}]{liang2015}%
  \BibitemOpen
  \bibfield  {author} {\bibinfo {author} {\bibfnamefont {E.}~\bibnamefont
  {Liang}}, \bibinfo {author} {\bibfnamefont {T.}~\bibnamefont {Clarke}},
  \bibinfo {author} {\bibfnamefont {A.}~\bibnamefont {Henderson}}, \bibinfo
  {author} {\bibfnamefont {W.}~\bibnamefont {Fu}}, \bibinfo {author}
  {\bibfnamefont {W.}~\bibnamefont {Lo}}, \bibinfo {author} {\bibfnamefont
  {D.}~\bibnamefont {Taylor}}, \bibinfo {author} {\bibfnamefont
  {P.}~\bibnamefont {Chaguine}}, \bibinfo {author} {\bibfnamefont
  {S.}~\bibnamefont {Zhou}}, \bibinfo {author} {\bibfnamefont {Y.}~\bibnamefont
  {Hua}}, \bibinfo {author} {\bibfnamefont {X.}~\bibnamefont {Cen}}, \bibinfo
  {author} {\bibfnamefont {X.}~\bibnamefont {Wang}}, \bibinfo {author}
  {\bibfnamefont {J.}~\bibnamefont {Kao}}, \bibinfo {author} {\bibfnamefont
  {H.}~\bibnamefont {Hasson}}, \bibinfo {author} {\bibfnamefont
  {G.}~\bibnamefont {Dyer}}, \bibinfo {author} {\bibfnamefont {K.}~\bibnamefont
  {Serratto}}, \bibinfo {author} {\bibfnamefont {N.}~\bibnamefont {Riley}},
  \bibinfo {author} {\bibfnamefont {M.}~\bibnamefont {Donovan}},\ and\ \bibinfo
  {author} {\bibfnamefont {T.}~\bibnamefont {Ditmire}},\ }\bibfield  {title}
  {\bibinfo {title} {High e+/e- ratio dense pair creation with {$10^{21}$}
  {W}/cm{$^2$} laser irradiating solid targets},\ }\href
  {https://doi.org/10.1038/srep13968} {\bibfield  {journal} {\bibinfo
  {journal} {Sci. Rep.}\ }\textbf {\bibinfo {volume} {5}},\ \bibinfo {pages}
  {13968} (\bibinfo {year} {2015})}\BibitemShut {NoStop}%
\bibitem [{\citenamefont {Xu}\ \emph {et~al.}(2016)\citenamefont {Xu},
  \citenamefont {Shen}, \citenamefont {Xu}, \citenamefont {Li}, \citenamefont
  {Yu}, \citenamefont {Li}, \citenamefont {Lu}, \citenamefont {Wang},
  \citenamefont {Wang}, \citenamefont {Liang}, \citenamefont {Leng},
  \citenamefont {Li},\ and\ \citenamefont {Xu}}]{xu2016}%
  \BibitemOpen
  \bibfield  {author} {\bibinfo {author} {\bibfnamefont {T.}~\bibnamefont
  {Xu}}, \bibinfo {author} {\bibfnamefont {B.}~\bibnamefont {Shen}}, \bibinfo
  {author} {\bibfnamefont {J.}~\bibnamefont {Xu}}, \bibinfo {author}
  {\bibfnamefont {S.}~\bibnamefont {Li}}, \bibinfo {author} {\bibfnamefont
  {Y.}~\bibnamefont {Yu}}, \bibinfo {author} {\bibfnamefont {J.}~\bibnamefont
  {Li}}, \bibinfo {author} {\bibfnamefont {X.}~\bibnamefont {Lu}}, \bibinfo
  {author} {\bibfnamefont {C.}~\bibnamefont {Wang}}, \bibinfo {author}
  {\bibfnamefont {X.}~\bibnamefont {Wang}}, \bibinfo {author} {\bibfnamefont
  {X.}~\bibnamefont {Liang}}, \bibinfo {author} {\bibfnamefont
  {Y.}~\bibnamefont {Leng}}, \bibinfo {author} {\bibfnamefont {R.}~\bibnamefont
  {Li}},\ and\ \bibinfo {author} {\bibfnamefont {Z.}~\bibnamefont {Xu}},\
  }\bibfield  {title} {\bibinfo {title} {Ultrashort megaelectronvolt positron
  beam generation based on laser-accelerated electrons},\ }\href
  {https://doi.org/10.1063/1.4943280} {\bibfield  {journal} {\bibinfo
  {journal} {Phys. Plasmas}\ }\textbf {\bibinfo {volume} {23}},\ \bibinfo
  {pages} {033109} (\bibinfo {year} {2016})}\BibitemShut {NoStop}%
\bibitem [{\citenamefont {Peebles}\ \emph {et~al.}(2021)\citenamefont
  {Peebles}, \citenamefont {Fiksel}, \citenamefont {Edwards}, \citenamefont
  {von~der Linden}, \citenamefont {Willingale}, \citenamefont {Mastrosimone},\
  and\ \citenamefont {Chen}}]{peebles2021}%
  \BibitemOpen
  \bibfield  {author} {\bibinfo {author} {\bibfnamefont {J.~L.}\ \bibnamefont
  {Peebles}}, \bibinfo {author} {\bibfnamefont {G.}~\bibnamefont {Fiksel}},
  \bibinfo {author} {\bibfnamefont {M.~R.}\ \bibnamefont {Edwards}}, \bibinfo
  {author} {\bibfnamefont {J.}~\bibnamefont {von~der Linden}}, \bibinfo
  {author} {\bibfnamefont {L.}~\bibnamefont {Willingale}}, \bibinfo {author}
  {\bibfnamefont {D.}~\bibnamefont {Mastrosimone}},\ and\ \bibinfo {author}
  {\bibfnamefont {H.}~\bibnamefont {Chen}},\ }\bibfield  {title} {\bibinfo
  {title} {Magnetically collimated relativistic charge-neutral
  electron–positron beams from high-power lasers},\ }\href
  {https://doi.org/10.1063/5.0053557} {\bibfield  {journal} {\bibinfo
  {journal} {Phys. Plasmas}\ }\textbf {\bibinfo {volume} {28}},\ \bibinfo
  {pages} {074501} (\bibinfo {year} {2021})}\BibitemShut {NoStop}%
\bibitem [{\citenamefont {Arrowsmith}\ \emph {et~al.}(2024)\citenamefont
  {Arrowsmith}, \citenamefont {Simon}, \citenamefont {Bilbao}, \citenamefont
  {Bott}, \citenamefont {Burger}, \citenamefont {Chen}, \citenamefont {Cruz},
  \citenamefont {Davenne}, \citenamefont {Efthymiopoulos}, \citenamefont
  {Froula}, \citenamefont {Goillot}, \citenamefont {Gudmundsson}, \citenamefont
  {Haberberger}, \citenamefont {Halliday}, \citenamefont {Hodge}, \citenamefont
  {Huffman}, \citenamefont {Iaquinta}, \citenamefont {Miniati}, \citenamefont
  {Reville}, \citenamefont {Sarkar}, \citenamefont {Schekochihin},
  \citenamefont {Silva}, \citenamefont {Simpson}, \citenamefont {Stergiou},
  \citenamefont {Trines}, \citenamefont {Vieu}, \citenamefont {Charitonidis},
  \citenamefont {Bingham},\ and\ \citenamefont {Gregori}}]{arrowsmith2024}%
  \BibitemOpen
  \bibfield  {author} {\bibinfo {author} {\bibfnamefont {C.~D.}\ \bibnamefont
  {Arrowsmith}}, \bibinfo {author} {\bibfnamefont {P.}~\bibnamefont {Simon}},
  \bibinfo {author} {\bibfnamefont {P.~J.}\ \bibnamefont {Bilbao}}, \bibinfo
  {author} {\bibfnamefont {A.~F.~A.}\ \bibnamefont {Bott}}, \bibinfo {author}
  {\bibfnamefont {S.}~\bibnamefont {Burger}}, \bibinfo {author} {\bibfnamefont
  {H.}~\bibnamefont {Chen}}, \bibinfo {author} {\bibfnamefont {F.~D.}\
  \bibnamefont {Cruz}}, \bibinfo {author} {\bibfnamefont {T.}~\bibnamefont
  {Davenne}}, \bibinfo {author} {\bibfnamefont {I.}~\bibnamefont
  {Efthymiopoulos}}, \bibinfo {author} {\bibfnamefont {D.~H.}\ \bibnamefont
  {Froula}}, \bibinfo {author} {\bibfnamefont {A.}~\bibnamefont {Goillot}},
  \bibinfo {author} {\bibfnamefont {J.~T.}\ \bibnamefont {Gudmundsson}},
  \bibinfo {author} {\bibfnamefont {D.}~\bibnamefont {Haberberger}}, \bibinfo
  {author} {\bibfnamefont {J.~W.~D.}\ \bibnamefont {Halliday}}, \bibinfo
  {author} {\bibfnamefont {T.}~\bibnamefont {Hodge}}, \bibinfo {author}
  {\bibfnamefont {B.~T.}\ \bibnamefont {Huffman}}, \bibinfo {author}
  {\bibfnamefont {S.}~\bibnamefont {Iaquinta}}, \bibinfo {author}
  {\bibfnamefont {F.}~\bibnamefont {Miniati}}, \bibinfo {author} {\bibfnamefont
  {B.}~\bibnamefont {Reville}}, \bibinfo {author} {\bibfnamefont
  {S.}~\bibnamefont {Sarkar}}, \bibinfo {author} {\bibfnamefont {A.~A.}\
  \bibnamefont {Schekochihin}}, \bibinfo {author} {\bibfnamefont {L.~O.}\
  \bibnamefont {Silva}}, \bibinfo {author} {\bibfnamefont {R.}~\bibnamefont
  {Simpson}}, \bibinfo {author} {\bibfnamefont {V.}~\bibnamefont {Stergiou}},
  \bibinfo {author} {\bibfnamefont {R.~M. G.~M.}\ \bibnamefont {Trines}},
  \bibinfo {author} {\bibfnamefont {T.}~\bibnamefont {Vieu}}, \bibinfo {author}
  {\bibfnamefont {N.}~\bibnamefont {Charitonidis}}, \bibinfo {author}
  {\bibfnamefont {R.}~\bibnamefont {Bingham}},\ and\ \bibinfo {author}
  {\bibfnamefont {G.}~\bibnamefont {Gregori}},\ }\bibfield  {title} {\bibinfo
  {title} {Laboratory realization of relativistic pair-plasma beams},\ }\href
  {https://doi.org/10.1038/s41467-024-49346-2} {\bibfield  {journal} {\bibinfo
  {journal} {Nat. Commun.}\ }\textbf {\bibinfo {volume} {15}},\ \bibinfo
  {pages} {5029} (\bibinfo {year} {2024})}\BibitemShut {NoStop}%
\bibitem [{\citenamefont {Streeter}\ \emph {et~al.}(2024)\citenamefont
  {Streeter}, \citenamefont {Colgan}, \citenamefont {Carderelli}, \citenamefont
  {Ma}, \citenamefont {Cavanagh}, \citenamefont {Los}, \citenamefont {Ahmed},
  \citenamefont {Antoine}, \citenamefont {Audet}, \citenamefont {Balcazar},
  \citenamefont {Calvin}, \citenamefont {Kettle}, \citenamefont {Mangles},
  \citenamefont {Najmudin}, \citenamefont {Rajeev}, \citenamefont {Symes},
  \citenamefont {Thomas},\ and\ \citenamefont {Sarri}}]{streeter2024}%
  \BibitemOpen
  \bibfield  {author} {\bibinfo {author} {\bibfnamefont {M.~J.~V.}\
  \bibnamefont {Streeter}}, \bibinfo {author} {\bibfnamefont {C.}~\bibnamefont
  {Colgan}}, \bibinfo {author} {\bibfnamefont {J.}~\bibnamefont {Carderelli}},
  \bibinfo {author} {\bibfnamefont {Y.}~\bibnamefont {Ma}}, \bibinfo {author}
  {\bibfnamefont {N.}~\bibnamefont {Cavanagh}}, \bibinfo {author}
  {\bibfnamefont {E.~E.}\ \bibnamefont {Los}}, \bibinfo {author} {\bibfnamefont
  {H.}~\bibnamefont {Ahmed}}, \bibinfo {author} {\bibfnamefont {A.~F.}\
  \bibnamefont {Antoine}}, \bibinfo {author} {\bibfnamefont {T.}~\bibnamefont
  {Audet}}, \bibinfo {author} {\bibfnamefont {M.~D.}\ \bibnamefont {Balcazar}},
  \bibinfo {author} {\bibfnamefont {L.}~\bibnamefont {Calvin}}, \bibinfo
  {author} {\bibfnamefont {B.}~\bibnamefont {Kettle}}, \bibinfo {author}
  {\bibfnamefont {S.~P.~D.}\ \bibnamefont {Mangles}}, \bibinfo {author}
  {\bibfnamefont {Z.}~\bibnamefont {Najmudin}}, \bibinfo {author}
  {\bibfnamefont {P.~P.}\ \bibnamefont {Rajeev}}, \bibinfo {author}
  {\bibfnamefont {D.~R.}\ \bibnamefont {Symes}}, \bibinfo {author}
  {\bibfnamefont {A.~G.~R.}\ \bibnamefont {Thomas}},\ and\ \bibinfo {author}
  {\bibfnamefont {G.}~\bibnamefont {Sarri}},\ }\bibfield  {title} {\bibinfo
  {title} {Narrow bandwidth, low-emittance positron beams from a
  laser-wakefield accelerator},\ }\href
  {https://doi.org/10.1038/s41598-024-56281-1} {\bibfield  {journal} {\bibinfo
  {journal} {Sci. Rep.}\ }\textbf {\bibinfo {volume} {14}},\ \bibinfo {pages}
  {6001} (\bibinfo {year} {2024})}\BibitemShut {NoStop}%
\bibitem [{\citenamefont {Noh}\ \emph {et~al.}(2024)\citenamefont {Noh},
  \citenamefont {Song}, \citenamefont {Mirzaie}, \citenamefont {Hojbota},
  \citenamefont {Kim}, \citenamefont {Lee}, \citenamefont {Won}, \citenamefont
  {Song}, \citenamefont {Song}, \citenamefont {Ryu}, \citenamefont {Nam},\ and\
  \citenamefont {Bang}}]{noh2024}%
  \BibitemOpen
  \bibfield  {author} {\bibinfo {author} {\bibfnamefont {Y.}~\bibnamefont
  {Noh}}, \bibinfo {author} {\bibfnamefont {J.}~\bibnamefont {Song}}, \bibinfo
  {author} {\bibfnamefont {M.}~\bibnamefont {Mirzaie}}, \bibinfo {author}
  {\bibfnamefont {C.~I.}\ \bibnamefont {Hojbota}}, \bibinfo {author}
  {\bibfnamefont {H.-i.}\ \bibnamefont {Kim}}, \bibinfo {author} {\bibfnamefont
  {S.}~\bibnamefont {Lee}}, \bibinfo {author} {\bibfnamefont {J.}~\bibnamefont
  {Won}}, \bibinfo {author} {\bibfnamefont {H.}~\bibnamefont {Song}}, \bibinfo
  {author} {\bibfnamefont {C.}~\bibnamefont {Song}}, \bibinfo {author}
  {\bibfnamefont {C.-M.}\ \bibnamefont {Ryu}}, \bibinfo {author} {\bibfnamefont
  {C.~H.}\ \bibnamefont {Nam}},\ and\ \bibinfo {author} {\bibfnamefont
  {W.}~\bibnamefont {Bang}},\ }\bibfield  {title} {\bibinfo {title}
  {Charge-neutral, gev-scale electron-positron pair beams produced using
  bremsstrahlung gamma rays},\ }\href
  {https://doi.org/10.1038/s42005-024-01527-7} {\bibfield  {journal} {\bibinfo
  {journal} {Commun. Phys.}\ }\textbf {\bibinfo {volume} {7}},\ \bibinfo
  {pages} {44} (\bibinfo {year} {2024})}\BibitemShut {NoStop}%
\bibitem [{\citenamefont {{Rohrlich}}(2007)}]{rohrlich2007}%
  \BibitemOpen
  \bibfield  {author} {\bibinfo {author} {\bibfnamefont {F.}~\bibnamefont
  {{Rohrlich}}},\ }\href {https://doi.org/10.1142/6220} {\emph {\bibinfo
  {title} {{Classical Charged Particles}}}},\ \bibinfo {edition} {3rd}\ ed.\
  (\bibinfo  {publisher} {World Scientific},\ \bibinfo {year}
  {2007})\BibitemShut {NoStop}%
\bibitem [{\citenamefont {Landau}\ and\ \citenamefont
  {Lifshitz}(1975)}]{landaulifshitz_vol2}%
  \BibitemOpen
  \bibfield  {author} {\bibinfo {author} {\bibfnamefont {L.~D.}\ \bibnamefont
  {Landau}}\ and\ \bibinfo {author} {\bibfnamefont {E.~M.}\ \bibnamefont
  {Lifshitz}},\ }\href@noop {} {\emph {\bibinfo {title} {The Classical Theory
  of Fields}}},\ \bibinfo {edition} {2nd}\ ed.\ (\bibinfo  {publisher}
  {Elsevier, Oxford},\ \bibinfo {year} {1975})\BibitemShut {NoStop}%
\bibitem [{\citenamefont {{Di Piazza}}\ \emph {et~al.}(2012)\citenamefont {{Di
  Piazza}}, \citenamefont {M\"uller}, \citenamefont {Hatsagortsyan},\ and\
  \citenamefont {Keitel}}]{dipiazza2012_review}%
  \BibitemOpen
  \bibfield  {author} {\bibinfo {author} {\bibfnamefont {A.}~\bibnamefont {{Di
  Piazza}}}, \bibinfo {author} {\bibfnamefont {C.}~\bibnamefont {M\"uller}},
  \bibinfo {author} {\bibfnamefont {K.~Z.}\ \bibnamefont {Hatsagortsyan}},\
  and\ \bibinfo {author} {\bibfnamefont {C.~H.}\ \bibnamefont {Keitel}},\
  }\bibfield  {title} {\bibinfo {title} {Extremely high-intensity laser
  interactions with fundamental quantum systems},\ }\href
  {https://doi.org/10.1103/RevModPhys.84.1177} {\bibfield  {journal} {\bibinfo
  {journal} {Rev. Mod. Phys.}\ }\textbf {\bibinfo {volume} {84}},\ \bibinfo
  {pages} {1177} (\bibinfo {year} {2012})}\BibitemShut {NoStop}%
\bibitem [{\citenamefont {Quin}\ \emph
  {et~al.}(2025{\natexlab{a}})\citenamefont {Quin}, \citenamefont {Di~Piazza},
  \citenamefont {Keitel},\ and\ \citenamefont {Tamburini}}]{quin2023}%
  \BibitemOpen
  \bibfield  {author} {\bibinfo {author} {\bibfnamefont {M.~J.}\ \bibnamefont
  {Quin}}, \bibinfo {author} {\bibfnamefont {A.}~\bibnamefont {Di~Piazza}},
  \bibinfo {author} {\bibfnamefont {C.~H.}\ \bibnamefont {Keitel}},\ and\
  \bibinfo {author} {\bibfnamefont {M.}~\bibnamefont {Tamburini}},\ }\bibfield
  {title} {\bibinfo {title} {Effect of interparticle fields and radiation
  reaction on beam dynamics},\ }\href
  {https://doi.org/10.1103/PhysRevResearch.7.023210} {\bibfield  {journal}
  {\bibinfo  {journal} {Phys. Rev. Res.}\ }\textbf {\bibinfo {volume} {7}},\
  \bibinfo {pages} {023210} (\bibinfo {year} {2025}{\natexlab{a}})}\BibitemShut
  {NoStop}%
\bibitem [{\citenamefont {Tamburini}\ \emph {et~al.}(2010)\citenamefont
  {Tamburini}, \citenamefont {Pegoraro}, \citenamefont {{Di Piazza}},
  \citenamefont {Keitel},\ and\ \citenamefont {Macchi}}]{tamburini2010}%
  \BibitemOpen
  \bibfield  {author} {\bibinfo {author} {\bibfnamefont {M.}~\bibnamefont
  {Tamburini}}, \bibinfo {author} {\bibfnamefont {F.}~\bibnamefont {Pegoraro}},
  \bibinfo {author} {\bibfnamefont {A.}~\bibnamefont {{Di Piazza}}}, \bibinfo
  {author} {\bibfnamefont {C.~H.}\ \bibnamefont {Keitel}},\ and\ \bibinfo
  {author} {\bibfnamefont {A.}~\bibnamefont {Macchi}},\ }\bibfield  {title}
  {\bibinfo {title} {Radiation reaction effects on radiation pressure
  acceleration},\ }\href {https://doi.org/10.1088/1367-2630/12/12/123005}
  {\bibfield  {journal} {\bibinfo  {journal} {New J. Phys.}\ }\textbf {\bibinfo
  {volume} {12}},\ \bibinfo {pages} {123005} (\bibinfo {year}
  {2010})}\BibitemShut {NoStop}%
\bibitem [{\citenamefont {Gonoskov}\ \emph {et~al.}(2022)\citenamefont
  {Gonoskov}, \citenamefont {Blackburn}, \citenamefont {Marklund},\ and\
  \citenamefont {Bulanov}}]{gonoskov2022}%
  \BibitemOpen
  \bibfield  {author} {\bibinfo {author} {\bibfnamefont {A.}~\bibnamefont
  {Gonoskov}}, \bibinfo {author} {\bibfnamefont {T.~G.}\ \bibnamefont
  {Blackburn}}, \bibinfo {author} {\bibfnamefont {M.}~\bibnamefont
  {Marklund}},\ and\ \bibinfo {author} {\bibfnamefont {S.~S.}\ \bibnamefont
  {Bulanov}},\ }\bibfield  {title} {\bibinfo {title} {Charged particle motion
  and radiation in strong electromagnetic fields},\ }\href
  {https://doi.org/10.1103/RevModPhys.94.045001} {\bibfield  {journal}
  {\bibinfo  {journal} {Rev. Mod. Phys.}\ }\textbf {\bibinfo {volume} {94}},\
  \bibinfo {pages} {045001} (\bibinfo {year} {2022})}\BibitemShut {NoStop}%
\bibitem [{\citenamefont {Fedotov}\ \emph {et~al.}(2023)\citenamefont
  {Fedotov}, \citenamefont {Ilderton}, \citenamefont {Karbstein}, \citenamefont
  {King}, \citenamefont {Seipt}, \citenamefont {Taya},\ and\ \citenamefont
  {Torgrimsson}}]{fedotov2023}%
  \BibitemOpen
  \bibfield  {author} {\bibinfo {author} {\bibfnamefont {A.}~\bibnamefont
  {Fedotov}}, \bibinfo {author} {\bibfnamefont {A.}~\bibnamefont {Ilderton}},
  \bibinfo {author} {\bibfnamefont {F.}~\bibnamefont {Karbstein}}, \bibinfo
  {author} {\bibfnamefont {B.}~\bibnamefont {King}}, \bibinfo {author}
  {\bibfnamefont {D.}~\bibnamefont {Seipt}}, \bibinfo {author} {\bibfnamefont
  {H.}~\bibnamefont {Taya}},\ and\ \bibinfo {author} {\bibfnamefont
  {G.}~\bibnamefont {Torgrimsson}},\ }\bibfield  {title} {\bibinfo {title}
  {Advances in qed with intense background fields},\ }\href
  {https://doi.org/10.1016/j.physrep.2023.01.003} {\bibfield  {journal}
  {\bibinfo  {journal} {Phys. Rep.}\ }\textbf {\bibinfo {volume} {1010}},\
  \bibinfo {pages} {1} (\bibinfo {year} {2023})}\BibitemShut {NoStop}%
\bibitem [{\citenamefont {Jackson}(1998)}]{jackson1998}%
  \BibitemOpen
  \bibfield  {author} {\bibinfo {author} {\bibfnamefont {J.~D.}\ \bibnamefont
  {Jackson}},\ }\href@noop {} {\emph {\bibinfo {title} {Classical
  Electrodynamics}}},\ \bibinfo {edition} {3rd}\ ed.\ (\bibinfo  {publisher}
  {John Wiley and Sons, Inc.},\ \bibinfo {year} {1998})\BibitemShut {NoStop}%
\bibitem [{\citenamefont {Winkler}\ \emph {et~al.}(2025)\citenamefont
  {Winkler}, \citenamefont {Trunk}, \citenamefont {H{\"u}bner}, \citenamefont
  {Martinez de~la Ossa}, \citenamefont {Jalas}, \citenamefont {Kirchen},
  \citenamefont {Agapov}, \citenamefont {Antipov}, \citenamefont {Brinkmann},
  \citenamefont {Eichner}, \citenamefont {Ferran~Pousa}, \citenamefont
  {H{\"u}lsenbusch}, \citenamefont {Palmer}, \citenamefont {Schnepp},
  \citenamefont {Schubert}, \citenamefont {Th{\'e}venet}, \citenamefont
  {Walker}, \citenamefont {Werle}, \citenamefont {Leemans},\ and\ \citenamefont
  {Maier}}]{winkler2025}%
  \BibitemOpen
  \bibfield  {author} {\bibinfo {author} {\bibfnamefont {P.}~\bibnamefont
  {Winkler}}, \bibinfo {author} {\bibfnamefont {M.}~\bibnamefont {Trunk}},
  \bibinfo {author} {\bibfnamefont {L.}~\bibnamefont {H{\"u}bner}}, \bibinfo
  {author} {\bibfnamefont {A.}~\bibnamefont {Martinez de~la Ossa}}, \bibinfo
  {author} {\bibfnamefont {S.}~\bibnamefont {Jalas}}, \bibinfo {author}
  {\bibfnamefont {M.}~\bibnamefont {Kirchen}}, \bibinfo {author} {\bibfnamefont
  {I.}~\bibnamefont {Agapov}}, \bibinfo {author} {\bibfnamefont {S.~A.}\
  \bibnamefont {Antipov}}, \bibinfo {author} {\bibfnamefont {R.}~\bibnamefont
  {Brinkmann}}, \bibinfo {author} {\bibfnamefont {T.}~\bibnamefont {Eichner}},
  \bibinfo {author} {\bibfnamefont {A.}~\bibnamefont {Ferran~Pousa}}, \bibinfo
  {author} {\bibfnamefont {T.}~\bibnamefont {H{\"u}lsenbusch}}, \bibinfo
  {author} {\bibfnamefont {G.}~\bibnamefont {Palmer}}, \bibinfo {author}
  {\bibfnamefont {M.}~\bibnamefont {Schnepp}}, \bibinfo {author} {\bibfnamefont
  {K.}~\bibnamefont {Schubert}}, \bibinfo {author} {\bibfnamefont
  {M.}~\bibnamefont {Th{\'e}venet}}, \bibinfo {author} {\bibfnamefont {P.~A.}\
  \bibnamefont {Walker}}, \bibinfo {author} {\bibfnamefont {C.}~\bibnamefont
  {Werle}}, \bibinfo {author} {\bibfnamefont {W.~P.}\ \bibnamefont {Leemans}},\
  and\ \bibinfo {author} {\bibfnamefont {A.~R.}\ \bibnamefont {Maier}},\
  }\bibfield  {title} {\bibinfo {title} {Active energy compression of a
  laser-plasma electron beam},\ }\href
  {https://doi.org/10.1038/s41586-025-08772-y} {\bibfield  {journal} {\bibinfo
  {journal} {Nature}\ }\textbf {\bibinfo {volume} {640}},\ \bibinfo {pages}
  {907} (\bibinfo {year} {2025})}\BibitemShut {NoStop}%
\bibitem [{\citenamefont {Quin}(2023)}]{quin2023_phd}%
  \BibitemOpen
  \bibfield  {author} {\bibinfo {author} {\bibfnamefont {M.~J.}\ \bibnamefont
  {Quin}},\ }\emph {\bibinfo {title} {Classical Radiation Reaction and
  Collective Behaviour}},\ \href {https://doi.org/10.11588/heidok.00034028}
  {Ph.D. thesis},\ \bibinfo  {school} {{University of Heidelberg}} (\bibinfo
  {year} {2023})\BibitemShut {NoStop}%
\bibitem [{\citenamefont {Wang}\ \emph {et~al.}(2017)\citenamefont {Wang},
  \citenamefont {Wang}, \citenamefont {Rockwood}, \citenamefont {Luther},
  \citenamefont {Hollinger}, \citenamefont {Curtis}, \citenamefont {Calvi},
  \citenamefont {Menoni},\ and\ \citenamefont {Rocca}}]{wang2017}%
  \BibitemOpen
  \bibfield  {author} {\bibinfo {author} {\bibfnamefont {Y.}~\bibnamefont
  {Wang}}, \bibinfo {author} {\bibfnamefont {S.}~\bibnamefont {Wang}}, \bibinfo
  {author} {\bibfnamefont {A.}~\bibnamefont {Rockwood}}, \bibinfo {author}
  {\bibfnamefont {B.~M.}\ \bibnamefont {Luther}}, \bibinfo {author}
  {\bibfnamefont {R.}~\bibnamefont {Hollinger}}, \bibinfo {author}
  {\bibfnamefont {A.}~\bibnamefont {Curtis}}, \bibinfo {author} {\bibfnamefont
  {C.}~\bibnamefont {Calvi}}, \bibinfo {author} {\bibfnamefont {C.~S.}\
  \bibnamefont {Menoni}},\ and\ \bibinfo {author} {\bibfnamefont {J.~J.}\
  \bibnamefont {Rocca}},\ }\bibfield  {title} {\bibinfo {title} {0.85 {PW}
  laser operation at 3.3 {Hz} and high-contrast ultrahigh-intensity $\lambda$ =
  400 {nm} second-harmonic beamline},\ }\href
  {https://doi.org/10.1364/OL.42.003828} {\bibfield  {journal} {\bibinfo
  {journal} {Opt. Lett.}\ }\textbf {\bibinfo {volume} {42}},\ \bibinfo {pages}
  {3828} (\bibinfo {year} {2017})}\BibitemShut {NoStop}%
\bibitem [{\citenamefont {Mironov}\ \emph {et~al.}(2011)\citenamefont
  {Mironov}, \citenamefont {Ginzburg}, \citenamefont {Lozhkarev}, \citenamefont
  {Luchinin}, \citenamefont {Kirsanov}, \citenamefont {Yakovlev}, \citenamefont
  {Khazanov},\ and\ \citenamefont {Shaykin}}]{mironov2011}%
  \BibitemOpen
  \bibfield  {author} {\bibinfo {author} {\bibfnamefont {S.~Y.}\ \bibnamefont
  {Mironov}}, \bibinfo {author} {\bibfnamefont {V.~N.}\ \bibnamefont
  {Ginzburg}}, \bibinfo {author} {\bibfnamefont {V.~V.}\ \bibnamefont
  {Lozhkarev}}, \bibinfo {author} {\bibfnamefont {G.~A.}\ \bibnamefont
  {Luchinin}}, \bibinfo {author} {\bibfnamefont {A.~V.}\ \bibnamefont
  {Kirsanov}}, \bibinfo {author} {\bibfnamefont {I.~V.}\ \bibnamefont
  {Yakovlev}}, \bibinfo {author} {\bibfnamefont {E.~A.}\ \bibnamefont
  {Khazanov}},\ and\ \bibinfo {author} {\bibfnamefont {A.~A.}\ \bibnamefont
  {Shaykin}},\ }\bibfield  {title} {\bibinfo {title} {Highly efficient
  second-harmonic generation of intense femtosecond pulses with a significant
  effect of cubic nonlinearity},\ }\href
  {https://doi.org/10.1070/QE2011v041n11ABEH014694} {\bibfield  {journal}
  {\bibinfo  {journal} {Quantum Electron.}\ }\textbf {\bibinfo {volume} {41}},\
  \bibinfo {pages} {963} (\bibinfo {year} {2011})}\BibitemShut {NoStop}%
\bibitem [{\citenamefont {Kim}\ \emph {et~al.}(2004)\citenamefont {Kim},
  \citenamefont {Kim}, \citenamefont {Baik}, \citenamefont {Umesh},\ and\
  \citenamefont {Nam}}]{kim2004}%
  \BibitemOpen
  \bibfield  {author} {\bibinfo {author} {\bibfnamefont {K.~T.}\ \bibnamefont
  {Kim}}, \bibinfo {author} {\bibfnamefont {C.~M.}\ \bibnamefont {Kim}},
  \bibinfo {author} {\bibfnamefont {M.-G.}\ \bibnamefont {Baik}}, \bibinfo
  {author} {\bibfnamefont {G.}~\bibnamefont {Umesh}},\ and\ \bibinfo {author}
  {\bibfnamefont {C.~H.}\ \bibnamefont {Nam}},\ }\bibfield  {title} {\bibinfo
  {title} {Single
  $\mathrm{sub}\text{\ensuremath{-}}50\text{\ensuremath{-}}\text{attosecond}$
  pulse generation from chirp-compensated harmonic radiation using material
  dispersion},\ }\href {https://doi.org/10.1103/PhysRevA.69.051805} {\bibfield
  {journal} {\bibinfo  {journal} {Phys. Rev. A}\ }\textbf {\bibinfo {volume}
  {69}},\ \bibinfo {pages} {051805} (\bibinfo {year} {2004})}\BibitemShut
  {NoStop}%
\bibitem [{\citenamefont {Liu}\ \emph {et~al.}(2023)\citenamefont {Liu},
  \citenamefont {Lu}, \citenamefont {Lu}, \citenamefont {Zhang}, \citenamefont
  {Wu}, \citenamefont {Wu}, \citenamefont {Lan}, \citenamefont {Zhang},
  \citenamefont {Lv}, \citenamefont {Ma}, \citenamefont {Xia}, \citenamefont
  {Wang}, \citenamefont {Cai}, \citenamefont {Zhao}, \citenamefont {Geng},
  \citenamefont {Xu},\ and\ \citenamefont {Yan}}]{liuPoP23}%
  \BibitemOpen
  \bibfield  {author} {\bibinfo {author} {\bibfnamefont {J.}~\bibnamefont
  {Liu}}, \bibinfo {author} {\bibfnamefont {H.}~\bibnamefont {Lu}}, \bibinfo
  {author} {\bibfnamefont {H.}~\bibnamefont {Lu}}, \bibinfo {author}
  {\bibfnamefont {H.}~\bibnamefont {Zhang}}, \bibinfo {author} {\bibfnamefont
  {X.}~\bibnamefont {Wu}}, \bibinfo {author} {\bibfnamefont {D.}~\bibnamefont
  {Wu}}, \bibinfo {author} {\bibfnamefont {H.}~\bibnamefont {Lan}}, \bibinfo
  {author} {\bibfnamefont {J.}~\bibnamefont {Zhang}}, \bibinfo {author}
  {\bibfnamefont {J.}~\bibnamefont {Lv}}, \bibinfo {author} {\bibfnamefont
  {Q.}~\bibnamefont {Ma}}, \bibinfo {author} {\bibfnamefont {Y.}~\bibnamefont
  {Xia}}, \bibinfo {author} {\bibfnamefont {Z.}~\bibnamefont {Wang}}, \bibinfo
  {author} {\bibfnamefont {J.}~\bibnamefont {Cai}}, \bibinfo {author}
  {\bibfnamefont {Y.}~\bibnamefont {Zhao}}, \bibinfo {author} {\bibfnamefont
  {Y.}~\bibnamefont {Geng}}, \bibinfo {author} {\bibfnamefont {X.}~\bibnamefont
  {Xu}},\ and\ \bibinfo {author} {\bibfnamefont {X.}~\bibnamefont {Yan}},\
  }\bibfield  {title} {\bibinfo {title} {Generation of $\sim$400 pc electron
  bunches in laser wakefield acceleration utilizing a structured plasma density
  profile},\ }\href {https://doi.org/10.1063/5.0161811} {\bibfield  {journal}
  {\bibinfo  {journal} {Phys. Plasmas}\ }\textbf {\bibinfo {volume} {30}},\
  \bibinfo {pages} {113103} (\bibinfo {year} {2023})}\BibitemShut {NoStop}%
\bibitem [{\citenamefont {Pellegrini}\ \emph {et~al.}(2016)\citenamefont
  {Pellegrini}, \citenamefont {Marinelli},\ and\ \citenamefont
  {Reiche}}]{pellegriniRMP16}%
  \BibitemOpen
  \bibfield  {author} {\bibinfo {author} {\bibfnamefont {C.}~\bibnamefont
  {Pellegrini}}, \bibinfo {author} {\bibfnamefont {A.}~\bibnamefont
  {Marinelli}},\ and\ \bibinfo {author} {\bibfnamefont {S.}~\bibnamefont
  {Reiche}},\ }\bibfield  {title} {\bibinfo {title} {The physics of x-ray
  free-electron lasers},\ }\href {https://doi.org/10.1103/RevModPhys.88.015006}
  {\bibfield  {journal} {\bibinfo  {journal} {Rev. Mod. Phys.}\ }\textbf
  {\bibinfo {volume} {88}},\ \bibinfo {pages} {015006} (\bibinfo {year}
  {2016})}\BibitemShut {NoStop}%
\bibitem [{\citenamefont {Emma}\ \emph {et~al.}(2025)\citenamefont {Emma},
  \citenamefont {Majernik}, \citenamefont {Swanson}, \citenamefont {Ariniello},
  \citenamefont {Gessner}, \citenamefont {Hessami}, \citenamefont {Hogan},
  \citenamefont {Knetsch}, \citenamefont {Larsen}, \citenamefont {Marinelli},
  \citenamefont {O'Shea}, \citenamefont {Perez}, \citenamefont {Rajkovic},
  \citenamefont {Robles}, \citenamefont {Storey},\ and\ \citenamefont
  {Yocky}}]{emmaPRL25}%
  \BibitemOpen
  \bibfield  {author} {\bibinfo {author} {\bibfnamefont {C.}~\bibnamefont
  {Emma}}, \bibinfo {author} {\bibfnamefont {N.}~\bibnamefont {Majernik}},
  \bibinfo {author} {\bibfnamefont {K.~K.}\ \bibnamefont {Swanson}}, \bibinfo
  {author} {\bibfnamefont {R.}~\bibnamefont {Ariniello}}, \bibinfo {author}
  {\bibfnamefont {S.}~\bibnamefont {Gessner}}, \bibinfo {author} {\bibfnamefont
  {R.}~\bibnamefont {Hessami}}, \bibinfo {author} {\bibfnamefont {M.~J.}\
  \bibnamefont {Hogan}}, \bibinfo {author} {\bibfnamefont {A.}~\bibnamefont
  {Knetsch}}, \bibinfo {author} {\bibfnamefont {K.~A.}\ \bibnamefont {Larsen}},
  \bibinfo {author} {\bibfnamefont {A.}~\bibnamefont {Marinelli}}, \bibinfo
  {author} {\bibfnamefont {B.}~\bibnamefont {O'Shea}}, \bibinfo {author}
  {\bibfnamefont {S.}~\bibnamefont {Perez}}, \bibinfo {author} {\bibfnamefont
  {I.}~\bibnamefont {Rajkovic}}, \bibinfo {author} {\bibfnamefont
  {R.}~\bibnamefont {Robles}}, \bibinfo {author} {\bibfnamefont
  {D.}~\bibnamefont {Storey}},\ and\ \bibinfo {author} {\bibfnamefont
  {G.}~\bibnamefont {Yocky}},\ }\bibfield  {title} {\bibinfo {title}
  {Experimental generation of extreme electron beams for advanced accelerator
  applications},\ }\href {https://doi.org/10.1103/PhysRevLett.134.085001}
  {\bibfield  {journal} {\bibinfo  {journal} {Phys. Rev. Lett.}\ }\textbf
  {\bibinfo {volume} {134}},\ \bibinfo {pages} {085001} (\bibinfo {year}
  {2025})}\BibitemShut {NoStop}%
\bibitem [{\citenamefont {Almeida}\ and\ \citenamefont
  {Vieira}(2023)}]{almeidal2023}%
  \BibitemOpen
  \bibfield  {author} {\bibinfo {author} {\bibfnamefont {R.~R.}\ \bibnamefont
  {Almeida}}\ and\ \bibinfo {author} {\bibfnamefont {J.}~\bibnamefont
  {Vieira}},\ }\href@noop {} {}\bibinfo {howpublished} {personal communication}
  (\bibinfo {year} {2023})\BibitemShut {NoStop}%
\bibitem [{\citenamefont {Fonseca}\ \emph {et~al.}(2002)\citenamefont
  {Fonseca}, \citenamefont {Silva}, \citenamefont {Tsung}, \citenamefont
  {Decyk}, \citenamefont {Lu}, \citenamefont {Ren}, \citenamefont {Mori},
  \citenamefont {Deng}, \citenamefont {Lee}, \citenamefont {Katsouleas},\ and\
  \citenamefont {Adam}}]{fonesca2002}%
  \BibitemOpen
  \bibfield  {author} {\bibinfo {author} {\bibfnamefont {R.~A.}\ \bibnamefont
  {Fonseca}}, \bibinfo {author} {\bibfnamefont {L.~O.}\ \bibnamefont {Silva}},
  \bibinfo {author} {\bibfnamefont {F.~S.}\ \bibnamefont {Tsung}}, \bibinfo
  {author} {\bibfnamefont {V.~K.}\ \bibnamefont {Decyk}}, \bibinfo {author}
  {\bibfnamefont {W.}~\bibnamefont {Lu}}, \bibinfo {author} {\bibfnamefont
  {C.}~\bibnamefont {Ren}}, \bibinfo {author} {\bibfnamefont {W.~B.}\
  \bibnamefont {Mori}}, \bibinfo {author} {\bibfnamefont {S.}~\bibnamefont
  {Deng}}, \bibinfo {author} {\bibfnamefont {S.}~\bibnamefont {Lee}}, \bibinfo
  {author} {\bibfnamefont {T.}~\bibnamefont {Katsouleas}},\ and\ \bibinfo
  {author} {\bibfnamefont {J.~C.}\ \bibnamefont {Adam}},\ }\bibfield  {title}
  {\bibinfo {title} {Osiris: A three-dimensional, fully relativistic particle
  in cell code for modeling plasma based accelerators},\ }in\ \href@noop {}
  {\emph {\bibinfo {booktitle} {Computational Science --- ICCS 2002}}}\
  (\bibinfo  {publisher} {Springer Berlin Heidelberg},\ \bibinfo {year}
  {2002})\ pp.\ \bibinfo {pages} {342--351}\BibitemShut {NoStop}%
\bibitem [{\citenamefont {Derouillat}\ \emph {et~al.}(2018)\citenamefont
  {Derouillat}, \citenamefont {Beck}, \citenamefont {Pérez}, \citenamefont
  {Vinci}, \citenamefont {Chiaramello}, \citenamefont {Grassi}, \citenamefont
  {Flé}, \citenamefont {Bouchard}, \citenamefont {Plotnikov}, \citenamefont
  {Aunai}, \citenamefont {Dargent}, \citenamefont {Riconda},\ and\
  \citenamefont {Grech}}]{derouillat2018}%
  \BibitemOpen
  \bibfield  {author} {\bibinfo {author} {\bibfnamefont {J.}~\bibnamefont
  {Derouillat}}, \bibinfo {author} {\bibfnamefont {A.}~\bibnamefont {Beck}},
  \bibinfo {author} {\bibfnamefont {F.}~\bibnamefont {Pérez}}, \bibinfo
  {author} {\bibfnamefont {T.}~\bibnamefont {Vinci}}, \bibinfo {author}
  {\bibfnamefont {M.}~\bibnamefont {Chiaramello}}, \bibinfo {author}
  {\bibfnamefont {A.}~\bibnamefont {Grassi}}, \bibinfo {author} {\bibfnamefont
  {M.}~\bibnamefont {Flé}}, \bibinfo {author} {\bibfnamefont {G.}~\bibnamefont
  {Bouchard}}, \bibinfo {author} {\bibfnamefont {I.}~\bibnamefont {Plotnikov}},
  \bibinfo {author} {\bibfnamefont {N.}~\bibnamefont {Aunai}}, \bibinfo
  {author} {\bibfnamefont {J.}~\bibnamefont {Dargent}}, \bibinfo {author}
  {\bibfnamefont {C.}~\bibnamefont {Riconda}},\ and\ \bibinfo {author}
  {\bibfnamefont {M.}~\bibnamefont {Grech}},\ }\bibfield  {title} {\bibinfo
  {title} {Smilei : A collaborative, open-source, multi-purpose
  particle-in-cell code for plasma simulation},\ }\href
  {https://doi.org/10.1016/j.cpc.2017.09.024} {\bibfield  {journal} {\bibinfo
  {journal} {Comput. Phys. Commun.}\ }\textbf {\bibinfo {volume} {222}},\
  \bibinfo {pages} {351} (\bibinfo {year} {2018})}\BibitemShut {NoStop}%
\bibitem [{\citenamefont {Pardal}\ \emph {et~al.}(2023)\citenamefont {Pardal},
  \citenamefont {Sainte-Marie}, \citenamefont {Reboul-Salze}, \citenamefont
  {Fonseca},\ and\ \citenamefont {Vieira}}]{pardal2023radio}%
  \BibitemOpen
  \bibfield  {author} {\bibinfo {author} {\bibfnamefont {M.}~\bibnamefont
  {Pardal}}, \bibinfo {author} {\bibfnamefont {A.}~\bibnamefont
  {Sainte-Marie}}, \bibinfo {author} {\bibfnamefont {A.}~\bibnamefont
  {Reboul-Salze}}, \bibinfo {author} {\bibfnamefont {R.~A.}\ \bibnamefont
  {Fonseca}},\ and\ \bibinfo {author} {\bibfnamefont {J.}~\bibnamefont
  {Vieira}},\ }\bibfield  {title} {\bibinfo {title} {Radio: An efficient
  spatiotemporal radiation diagnostic for particle-in-cell codes},\ }\href
  {https://doi.org/10.1016/j.cpc.2022.108634} {\bibfield  {journal} {\bibinfo
  {journal} {Comput. Phys. Commun.}\ }\textbf {\bibinfo {volume} {285}},\
  \bibinfo {pages} {108634} (\bibinfo {year} {2023})}\BibitemShut {NoStop}%
\bibitem [{\citenamefont {Salamin}\ \emph {et~al.}(2002)\citenamefont
  {Salamin}, \citenamefont {Mocken},\ and\ \citenamefont
  {Keitel}}]{salamin2002}%
  \BibitemOpen
  \bibfield  {author} {\bibinfo {author} {\bibfnamefont {Y.~I.}\ \bibnamefont
  {Salamin}}, \bibinfo {author} {\bibfnamefont {G.~R.}\ \bibnamefont
  {Mocken}},\ and\ \bibinfo {author} {\bibfnamefont {C.~H.}\ \bibnamefont
  {Keitel}},\ }\bibfield  {title} {\bibinfo {title} {Electron scattering and
  acceleration by a tightly focused laser beam},\ }\href
  {https://doi.org/10.1103/PhysRevSTAB.5.101301} {\bibfield  {journal}
  {\bibinfo  {journal} {Phys. Rev. ST Accel. Beams}\ }\textbf {\bibinfo
  {volume} {5}},\ \bibinfo {pages} {101301} (\bibinfo {year}
  {2002})}\BibitemShut {NoStop}%
\bibitem [{\citenamefont {Abramowitz}\ and\ \citenamefont
  {Stegun}(1968)}]{abramowitz1968}%
  \BibitemOpen
  \bibfield  {author} {\bibinfo {author} {\bibfnamefont {M.}~\bibnamefont
  {Abramowitz}}\ and\ \bibinfo {author} {\bibfnamefont {I.}~\bibnamefont
  {Stegun}},\ }\href@noop {} {\emph {\bibinfo {title} {Handbook of Mathematical
  Functions with Formulas, Graphs, and Mathematical Tables}}}\ (\bibinfo
  {publisher} {U.S. Government Printing Office},\ \bibinfo {year}
  {1968})\BibitemShut {NoStop}%
\bibitem [{\citenamefont {Quin}\ \emph
  {et~al.}(2025{\natexlab{b}})\citenamefont {Quin}, \citenamefont {Di~Piazza},
  \citenamefont {Erciyes}, \citenamefont {Keitel},\ and\ \citenamefont
  {Tamburini}}]{figuredata}%
  \BibitemOpen
  \bibfield  {author} {\bibinfo {author} {\bibfnamefont {M.~J.}\ \bibnamefont
  {Quin}}, \bibinfo {author} {\bibfnamefont {A.}~\bibnamefont {Di~Piazza}},
  \bibinfo {author} {\bibfnamefont {C.}~\bibnamefont {Erciyes}}, \bibinfo
  {author} {\bibfnamefont {C.~H.}\ \bibnamefont {Keitel}},\ and\ \bibinfo
  {author} {\bibfnamefont {M.}~\bibnamefont {Tamburini}},\ }\href@noop {}
  {}\bibinfo {howpublished} {Data for the figures in this manuscript:
  \url{https://doi.org/10.5281/zenodo.15696372}} (\bibinfo {year}
  {2025}{\natexlab{b}})\BibitemShut {NoStop}%
\end{thebibliography}%

\sectionmajor{Acknowledgements}
\smallskip

\noindent
This article comprises part of the PhD work of Michael J. Quin, which was successfully defended at Heidelberg University on the 18th of October, 2023. The authors wish to thank Rafael R. Almeida, Gianluca Aldo Geloni, Anna Golinelli, Christian Ott, Thomas Pfeifer, Brian Reville, Gianluca Sarri, and Jorge Vieira for helpful discussions.

This material is based upon work supported by the U.S. Department of Energy [National Nuclear Security Administration] University of Rochester ``National Inertial Confinement Fusion Program'' under Award Number DE-NA0004144 and U.S. Department of Energy, Office of Science, under Award Number DE-SC0021057.

This report was prepared as an account of work sponsored by an agency of the United States Government. Neither the United States Government nor any agency thereof, nor any of their employees, makes any warranty, express or implied, or assumes any legal liability or responsibility for the accuracy, completeness, or usefulness of any information, apparatus, product, or process disclosed, or represents that its use would not infringe privately owned rights. Reference herein to any specific commercial product, process, or service by trade name, trademark, manufacturer, or otherwise does not necessarily constitute or imply its endorsement, recommendation, or favoring by the United States Government or any agency thereof. The views and opinions of authors expressed herein do not necessarily state or reflect those of the United States Government or any agency thereof.%\\

\sectionmajor{Author Contributions}
\smallskip

\noindent
The idea for this paper originated with M.J.Q., A.D.P., and M.T. Point-particle code development, simulations, and analysis were performed by M.J.Q., with oversight by M.T. PIC code development and simulations were performed by C.E., with oversight by M.T. Interpretation and theoretical work were carried out by M.J.Q., A.D.P., and M.T., with oversight by C.H.K. The manuscript was written by M.J.Q., C.E., A.D.P., and M.T., with contributions from C.H.K.%\\

\sectionmajor{Competing interests}
\smallskip

\noindent
The authors declare no competing interests.

\end{document}